\documentclass[pre,groupedaddress,superscriptaddress,reprint,twocolumn,showpacs,notitlepage]{revtex4}

\usepackage{amsfonts}
\usepackage{amsmath}
\usepackage{amssymb}
\usepackage{graphicx}
\usepackage{hyperref}
\usepackage{color}
\usepackage{mathrsfs}
\usepackage{bbm}
\usepackage{subfigure}
\usepackage{times,txfonts}

\newcommand{\ignore}[1]{}
\newcommand{\nobibentry}[1]{{\let\nocite\ignore\bibentry{#1}}}
\newcommand{\bibfnamefont}[1]{#1}
\newcommand{\bibnamefont}[1]{#1}

\newcommand{\ket}[1]{\left\vert#1\right\rangle}
\newcommand{\bra}[1]{\left\langle#1\right\vert}
\newcommand{\eket}[2]{\left\vert#1_{{#2}}\right\rangle}
\newcommand{\ebra}[2]{\left\langle#1_{{#2}}\right\vert}
\newcommand{\eeket}[4]{\left\vert #1_{{#3}} #2_{{#4}}\right\rangle}
\newcommand{\eebra}[4]{\left\langle #1_{{#3}} #2_{{#4}} \right\vert}
\newcommand{\eeeket}[6]{\left\vert #1_{{#4}} #2_{{#5}} #3_{{#6}} \right\rangle}
\newcommand{\eeebra}[6]{\left\langle #1_{{#4}} #2_{{#5}} #3_{{#6}} \right\vert}

\begin{document}
%\nobibliography*

\title{Performance bound for quantum absorption refrigerators}

\author{Luis A. Correa}
\email{lacorrea@ull.es}
\affiliation{School of Mathematical Sciences, The University of Nottingham, University Park, Nottingham NG7 2RD, UK}
\affiliation{IUdEA Instituto Universitario de Estudios Avanzados, Universidad de La Laguna, 38203 Spain}
\affiliation{Dpto. F\'{\i}sica Fundamental, Experimental, Electr\'{o}nica y
Sistemas, Universidad de La Laguna, 38203 Spain}

\author{Jos\'{e} P. Palao}
\affiliation{IUdEA Instituto Universitario de Estudios Avanzados, Universidad de La Laguna, 38203 Spain}
\affiliation{Departamento de F\'{i}sica Fundamental II, Universidad de La Laguna, La Laguna 38204, Spain}

\author{Gerardo Adesso}
\affiliation{School of Mathematical Sciences, The University of Nottingham, University Park, Nottingham NG7 2RD, UK}

\author{Daniel Alonso}
\affiliation{IUdEA Instituto Universitario de Estudios Avanzados, Universidad de La Laguna, 38203 Spain}
\affiliation{Dpto. F\'{\i}sica Fundamental, Experimental, Electr\'{o}nica y
Sistemas, Universidad de La Laguna, 38203 Spain}

\date{April 04, 2013}

\begin{abstract}
An implementation of quantum absorption chillers with three qubits has been recently proposed, that is ideally able to reach the Carnot performance regime. Here we study the working efficiency of such self-contained refrigerators, adopting a consistent treatment of dissipation effects. We demonstrate that the coefficient of performance at maximum cooling power is upper bounded by $3/4$ of the Carnot performance. The result is independent of the details of the system and the equilibrium temperatures of the external baths. We provide design prescriptions that saturate the bound in the limit of a large difference between the operating temperatures.  Our study suggests that delocalized dissipation,  which must be taken into account for a proper modelling of the machine-baths interaction, is a fundamental source of irreversibility which prevents the refrigerator from approaching the Carnot performance arbitrarily closely in practice. The potential role of quantum correlations in the operation of these machines is also investigated.
\end{abstract}

\maketitle

\section{Introduction}
The study of quantum thermal machines has attracted an increasing attention over the last years. This is motivated on one hand by the fundamental interest in understanding the emergence of basic thermodynamical principles at the quantum mechanical level \cite{geva1996quantum,PhysRevE.64.056130,PhysRevLett.108.070604,PhysRevE.85.061126,PhysRevLett.105.130401,1751-8121-44-49-492002,PhysRevE.85.051117,1209.1190,PhysRevE.82.011120,natpop,guld,renner}, and on the other hand, by the potential technological applications of these machines, for instance to control the heat transport in nanoengineered devices \cite{giazotto,0295-5075_97_4_40003,1210.3649,PhysRevLett.109.203006}. In particular, several models have been proposed \cite{geva1996quantum,PhysRevE.64.056130,PhysRevLett.108.070604,PhysRevE.85.061126,PhysRevLett.105.130401,1751-8121-44-49-492002,PhysRevE.85.051117} realizing {\it quantum absorption chillers}, that is, refrigerators in which the external source of work is replaced by a heat bath. A realization of such refrigerators, which has been introduced in \cite{PhysRevLett.105.130401,1751-8121-44-49-492002,PhysRevE.85.051117}, consists of three interacting qubits, with a vanishingly small interaction strength, each one in contact with a heat bath. In spite of the technological challenges behind its physical implementation, this machine can be experimentally realized, e.g., with superconducting qubits or arrays of quantum dots \cite{0295-5075_97_4_40003, 1210.3649}. Furthermore, its operation may be understood in a very neat way,  providing a  physical insight into the sources of irreversibility in absorption chillers \cite{PhysRevE.85.051117}.

It has been predicted that, with a suitable choice of the machine parameters, such a refrigerator can ideally attain a coefficient of performance (COP) reaching the Carnot bound  $\varepsilon_C$ \cite{1751-8121-44-49-492002}. However, we argue that the central assumption of vanishing mutual interaction between the refrigerator qubits cannot be realistically maintained. As long as the interaction is finite, each bath will exchange energy with the whole three-qubit system, rather than just locally with the single qubit to which it is connected.  This is usually the case with any interacting multiparticle dissipative system. As we shall show, the resulting delocalized dissipation prevents the refrigerator from approaching the Carnot limit arbitrarily closely, thus embodying a fundamental source of irreversibility that is expected to arise in all concrete implementations.

This situation is reminiscent to that of realistic heat engines or cooling cycles, topical areas of study in {\it finite-time thermodynamics}. There, the finite heat transfer rates constitute the essential source of irreversibility which makes the Carnot bound unattainable in practice. For this reason, an important line of research deals with devising alternative, tight performance bounds, such that some suitable figure of merit of the machine under consideration is maximized \cite{curzon1975efficiency,PhysRevLett.78.3241,PhysRevLett.105.150603,PhysRevE.86.011127,PhysRevE.81.051129}. In this spirit, we address the following question: What is the highest achievable {\it COP at maximum cooling power} for the quantum absorption refrigerators of Refs.~\cite{PhysRevLett.105.130401,1751-8121-44-49-492002,PhysRevE.85.051117}? Answering this question would provide a practical performance bound against which the efficiency of any future realization of these machines could be benchmarked.

Another relevant and related question to ask is whether the ``quantumness'' of the refrigerator, as revealed for instance by the stationary quantum discord \cite{olliver20011,henderson20011}, plays any role in its operation. This would help to unveil connections between quantum correlations and efficient energy transport out of equilibrium, that so far have remained elusive.

In this paper we answer both questions. In the first place, by considering unstructured bosonic baths and a consistent dissipative qubit-bath interaction, we find that the COP at maximum power is tightly upper bounded by $\frac34 \varepsilon_C$, where $\varepsilon_C=\left(1-\frac{T_h}{T_w}\right)\big/\left(\frac{T_h}{T_c}-1\right)$ is the Carnot COP and $\{T_w,\,T_h,\,T_w\}$ are the three temperatures between which the refrigerator operates. We also give sufficient conditions to saturate this bound in the limit of large temperature difference $T_c/T_h\ll 1$. Secondly, we issue a comprehensive analysis of stationary bipartite quantum correlations in the various relevant qubit-qubit partitions. Although a nonvanishing discord is always found in a specific partition, it does not relate with the stationary heat flows, reinforcing the idea that this family of thermal machines operates in an effectively classical way \cite{PhysRevE.85.051117}, despite having a genuinely quantum physical support.

The manuscript is organized as follows:
In Section~\ref{secPrelim} we introduce the microscopic model of the three-qubit refrigerator object of our study. In Section~\ref{secDeloc} we address its reduced dynamics via a Lindblad-type master equation (whose derivation is deferred to Appendix~\ref{secAppA}), that allows for a \emph{consistent} treatment of dissipation. We also point how the \emph{delocalized dissipation} effects, unavoidable in practice, prevent the refrigerator from being maximally efficient. In Section~\ref{secBound} we demonstrate the existence of general upper bounds on the coefficient of performance of the refrigerator at maximum cooling power, and provide design prescriptions to saturate such bounds (a supporting analytical proof is presented in Appendix~\ref{secAppB}). In Section~\ref{secQC} we report our complete study  of stationary bipartite quantum correlations in the system. Finally, we draw our conclusions in Section~\ref{secConcl}.

\section{Preliminaries}\label{secPrelim}

\begin{figure}[t]
\includegraphics[width=8cm]{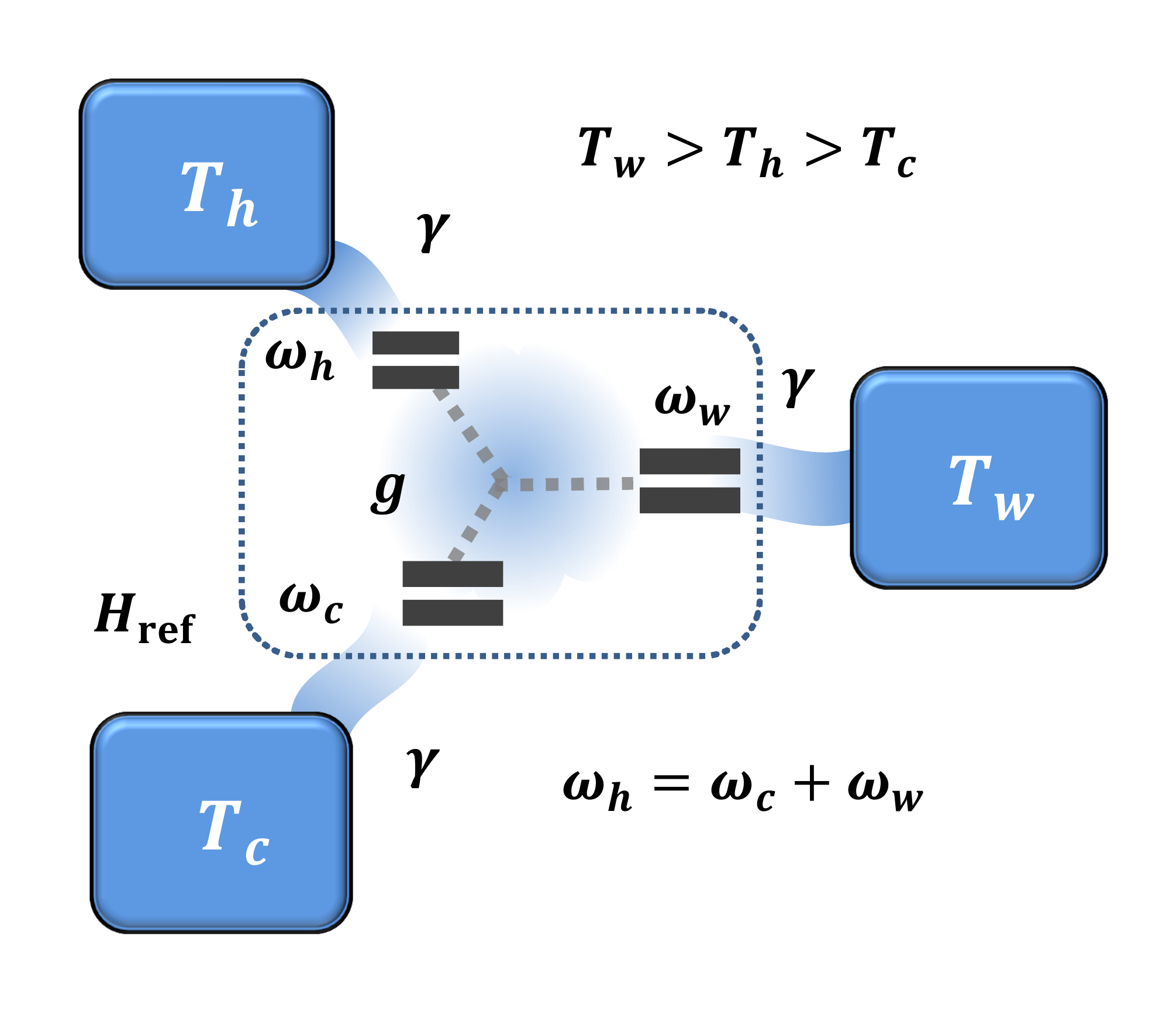}
\caption{(Color online). Schematic representation of the three-qubit absorption refrigerator. The refrigerator qubits dissipate into their respective baths, with equilibrium temperatures $T_w>T_h>T_c$, at a rate $\gamma$. The three-body interaction, of strength $g$, allows for energy exchange between the refrigerator qubits, whose energy spacings $\omega_\alpha$ are required to satisfy $\omega_h=\omega_c+\omega_w$.}
\label{fig1}
\end{figure}

Let us begin by introducing the total Hamiltonian of the refrigerator. The Hilbert space of the system is $\mathcal{H}_{S} = \mathcal{H}_{S,w} \otimes \mathcal{H}_{S,h} \otimes \mathcal{H}_{S,c}$, where we label the three qubits as `work', `hot', and `cold' ($w$, $h$, $c$), after the heat baths to which each of them is connected (see Fig.~\ref{fig1}). Their free Hamiltonians are
\begin{equation}
H_{0,\alpha}=\omega_\alpha\eket{1}{\alpha}\ebra{1}{\alpha},
\end{equation}
where $\alpha=\{w,\,h,\,c\}$, and we work in natural units $\hbar=k_B=1$. The corresponding bosonic baths are given by
\begin{equation}
H_{B,\alpha}=\sum\nolimits_{\lambda}\omega_{\lambda}a_{\alpha,\lambda}^{\dagger}a_{\alpha,\lambda}.
\end{equation}
As local qubit-bath dissipative coupling, we choose terms of the form
\begin{equation}
H_{D,\alpha}=\sqrt{\gamma}\big(c_{x_{\alpha}} \sigma_{x_{\alpha}} + c_{y_{\alpha}} \sigma_{y_{\alpha}}\big)\otimes\sum\nolimits_{\lambda}g_{\lambda}\big(a_{\alpha,\lambda}-a_{\alpha,\lambda}^{\dagger}\big),
\end{equation}
where $\{c_{x_{\alpha}},\,c_{y_{\alpha}}\}\in\mathbbm{R}$, and $g_{\lambda}\propto\sqrt{\omega_{\lambda}}$ to ensure flat spectral densities $J\left(\omega\right)\sim g^2_{\alpha,\lambda}/\omega_{\alpha,\lambda}$ \cite{breuer2002theory}. Here, we absorbed the order of magnitude of $J\left(\omega\right)$ into the dissipation rate $\gamma$. With no loss of generality, we can set $c_{x_{\alpha}}=1$ and $c_{y_{\alpha}}=0$. This kind of system-environment coupling stands e.g. for the dipole interaction between a two-level atom and the electromagnetic field at thermal equilibrium \cite{breuer2002theory}.

It only remains to specify the three body interaction between the qubits, which in our case is
\begin{equation}
H_{I}=g\left(\eeeket{1}{0}{1}{w}{h}{c}\eeebra{0}{1}{0}{w}{h}{c} + \, \text{h.c.}\right),
\end{equation}
where $g$ is the interaction strength. The qubit energies are chosen as $\omega_h=\omega_c+\omega_w$ ($\omega_h>\omega_w$), so that the subspace $\{\eeeket{1}{0}{1}{w}{h}{c},\,\eeeket{0}{1}{0}{w}{h}{c}\}$ is approximately degenerate, as long as $g\ll 1$.

The total Hamiltonian is then finally
\begin{equation}
\label{serrano}
H_T=\sum\nolimits_{\alpha}H_{0,\alpha}+H_I+\sum\nolimits_{\alpha}H_{D,\alpha}+\sum\nolimits_{\alpha}H_{B,\alpha}
\,.
\end{equation}

We now briefly explain how the operation of the refrigerator may be understood (see \cite{PhysRevE.85.051117} for details). In the ideal scenario of vanishing $g$, the reduced stationary state of the work and hot qubits $\varrho^{\infty}_{w,h}\equiv\text{Tr}_{c}\,\varrho$, has high fidelity with $\tau_{w}\otimes\tau_{h}$, where $\tau_{\alpha}=Z_{\alpha}^{-1}e^{-H_{0,\alpha}/T_{\alpha}}$ stands for the thermal state of qubit $\alpha$ at the equilibrium temperature $T_{\alpha}$, and $\varrho$ denotes the reduced state of the three qubits after tracing out the heat baths.

It follows that the truncation of $\varrho^{\infty}_{w,h}$ into the two-dimensional subspace $\mathcal{H}_{S,v}$ of $\mathcal{H}_{S,w}\otimes\mathcal{H}_{S,h}$ spanned by $\{\eeket{1}{0}{w}{h},\eeket{0}{1}{w}{h}\}$, which defines a `machine virtual qubit' $v$, has an effective virtual temperature approximately given by
\begin{equation*}
T_v\equiv \frac{\omega_h-\omega_w}{\omega_h/T_h-\omega_w/T_w}.
\end{equation*}

The interaction $H_I$ allows the cold qubit to exchange energy with the machine virtual qubit, while being simultaneously thermalized by the cold bath through $H_{D,c}$. Suitable choice of frequencies and temperatures may result in $T_v<T_c$, so that the (non-equilibrium) stationary state $\varrho^{\infty}_c$ is effectively \emph{colder} than $T_c$. The excited state population deficit in $\varrho^{\infty}_c$ is compensated by a net energy transfer from the cold bath (that stands for the object to cool) into the refrigerator. This is what we shall understand by \emph{cooling}. The machine therefore just mediates between the cold object at temperature $T_c$ and a suitably filtered virtual temperature $T_v$. Thermalization is then responsible for cooling within the \emph{cooling window} $0\leq T_v \leq T_c$, or in terms of the cold frequency $\omega_c$
\begin{equation}
0 < \omega_c < \frac{\big(T_w-T_h\big)T_c}{\big(T_w-T_c\big)T_h}\omega_h
\label{cwindow}\end{equation}
When $T_v=T_c$ and always under the assumption of localized dissipation, which is consistently realized only for vanishing $g$, the refrigerator would in principle saturate the Carnot bound $\varepsilon_C$ on COP \cite{1751-8121-44-49-492002}.

\section{Realistic modelling of the dissipation}\label{secDeloc}
\subsection{The quantum master equation}
We shall now extend the model of Refs.~\cite{PhysRevLett.105.130401,1751-8121-44-49-492002,PhysRevE.85.051117} to consistently account for the dissipative dynamics of the refrigerator in a the realistic scenario of even very small but nonvanishing coupling strength $g$.

We can derive a general equation of motion for the qubits from first principles, by employing the standard Born-Markov assumption of weak-memoryless system-environment interaction \cite{breuer2002theory}. Such master equation, whose complete derivation is reported in Appendix~\ref{secAppA}, is written as
\begin{multline}
\dot{\varrho}=-i\left[H_\text{ref},\varrho\right]+\sum_{\alpha}\mathcal{D}_{\alpha}\left[\varrho\right]\\=-i\left[H_\text{ref},\varrho\right]+\sum_{\alpha,\omega} \Gamma_{\alpha,\omega} \left( A_{\alpha,\omega}\varrho A^{\dagger}_{\alpha,\omega} -\mbox{$\frac{1}{2}$} \{ A^{\dagger}_{\alpha,\omega} A_{\alpha,\omega}, \varrho \}_+\right),
\label{master}\end{multline}
where $H_{\text{ref}}\equiv\sum_{\alpha}H_{0,\alpha}+H_I$. The spectral correlation tensor, denoted by $\Gamma_{\alpha,\omega}$, is proportional to the real part of the power spectra of the bath correlation functions. Note that since the heat baths are independent, the correlation tensor is diagonal in $\alpha$. We use its explicit form for electromagnetic radiation at thermal equilibrium:  $\Gamma_{\alpha,\omega}\propto \omega^3 \exp{(\omega/2T_\alpha)} \left(\sinh{\omega/2T_\alpha}\right)^{-1}$ \cite{breuer2002theory}.

The non-Hermitian Lindblad operator $A_{\alpha,\omega}$, associated with bath $\alpha$, performs transitions of frequency $\omega$ at rate $\Gamma_{\alpha,\omega}$, between the ($g$-independent) eigenstates of the refrigerator Hamiltonian $H_{\text{ref}}$. They result from the decomposition of the system-environment couplings ($\sqrt{\gamma}\sigma_{x_{\alpha}}=\sum_{\omega}A_{\alpha,\omega}$) as eigen-operators of the refrigerator Hamiltonian ($\left[H_{\text{ref}},\,A_{\alpha,\omega}\right]=-\omega A_{\alpha,\omega}$). Note that the corrections to $H_\text{ref}$ resulting from the system-environment interaction (i.e. Lamb shift Hamiltonian) have been neglected in Eq.~\eqref{master}, and that the rotating wave approximation is implicit its derivation. Therefore the time scale of the system $\tau_S\sim\max{ \lbrace g^{-1},\,\omega^{-1}_{\alpha}\rbrace }$ must be much smaller than the dissipation time, i.e. $\tau_S \ll \gamma^{-1}$.

Our dissipative system-environment coupling operators $\sqrt{\gamma}\sigma_{x_{\alpha}}$ give rise to six open decay channels (for each $\alpha$), associated with the frequencies $\{\pm \omega_\alpha,\,\pm\omega_{\alpha} \pm g\}$. Consider, for instance, the Lindblad operators within the cold \emph{dissipator} $\mathcal{D}_{c}$: While the operators $A_{c,\pm\omega_{c}}$ produce transitions
\begin{equation*}
\eeeket{0}{0}{0}{w}{h}{c} \leftrightarrow \eeeket{0}{0}{1}{w}{h}{c},\,\eeeket{1}{1}{0}{w}{h}{c} \leftrightarrow \eeeket{1}{1}{1}{w}{h}{c},
\end{equation*}
in which the cold bath exchanges energy \emph{locally} with the cold qubit only, the remaining operators, e.g. $A_{c,\pm\omega_{c}+g}$, are instead responsible for processes like
\begin{align*}
\eeeket{1}{0}{0}{w}{h}{c} &\leftrightarrow  \left(\eeeket{1}{0}{1}{w}{h}{c}\pm\eeeket{0}{1}{0}{w}{h}{c}\right)/\sqrt{2}\\
\eeeket{0}{1}{1}{w}{h}{c} &\leftrightarrow  \left(\eeeket{1}{0}{1}{w}{h}{c}\mp\eeeket{0}{1}{0}{w}{h}{c}\right)/\sqrt{2},
\end{align*}
in which bath $c$ now exchanges energy with the work and hot qubit as well, in a \emph{delocalized} way. It is in this sense, that we refer to $\mathcal{D}_{\alpha}$ as modelling a {\it delocalized dissipation} effect.

Of course, as the limit of vanishing coupling $g$ is approached, the rates $\Gamma_{\alpha,\pm \omega + g}$ and $\Gamma_{\alpha,\pm \omega - g}$ become equal, and all delocalized transitions tend to compensate each other. For $g=0$, only two (local) decay channels remain open for each bath, namely $A_{\alpha,\pm \omega_{\alpha}}(0)\propto\sigma_{\alpha,\mp}\otimes\mathbbm{1}_{\overline{\alpha}}$, which stands for the usual ladder operators for qubit $\alpha$ (the remaining qubits are denoted as $\overline{\alpha}$).  The idealized model of Refs.~\cite{PhysRevLett.105.130401,1751-8121-44-49-492002,PhysRevE.85.051117} is recovered in this limit. Note that since the frequencies appear exponentiated in the spectral correlation tensor, delocalized dissipation effects are intuitively expected to be still relevant (i.e. $\Gamma_{\pm\omega_{\alpha}+g}\not\simeq\Gamma_{\pm\omega_{\alpha}-g}$) even for arbitrarily small $g\lll 1$ (see Appendix~\ref{secAppA}).

Equipped with the stationary solutions of the Markovian master equation Eq.~\eqref{master}, we can compute the central quantities of our study, namely the rates at which energy from each bath is fed into the system, i.e., the heat currents. These are given as $\dot{\mathcal{Q}}_{\alpha}=\text{Tr}\{H_{\text{ref}} \mathcal{D}_{\alpha}\left[\varrho^{\infty}\right]\}$ \cite{breuer2002theory}. In particular, we refer to $\dot{\mathcal{Q}}_c$ as \emph{cooling power}. Therefore, the COP reads: $\varepsilon=\dot{\mathcal{Q}}_c/\dot{\mathcal{Q}}_w$ \cite{PhysRevE.64.056130,1751-8121-44-49-492002}.

\subsection{Delocalized dissipation and irreversibility}

Let now us comment on our intuition linking delocalized dissipation with \emph{irreversibility} in the operation of the machine. The Carnot COP $\varepsilon_C$ is realized at the upper limit of the cooling window Eq.~\eqref{cwindow}, whenever $\varepsilon=\dot{\mathcal{Q}}_c/\dot{\mathcal{Q}}_w=\omega_c/\omega_w$ \cite{1751-8121-44-49-492002}. This would be the case (for arbitrary $g$) if one replaced the consistent Eq.~\eqref{master} with a ``localized'' master equation such as:
\begin{equation}
\mbox{$\dot{\varrho}=-i\left[H_\text{ref},\varrho\right]+\sum_{\alpha}\mathscr{D}_{\alpha}\otimes\mathbbm{1}_{\overline{\alpha}}\left[\varrho\right]$},
\label{masterlocal}\end{equation}
like the one used in \cite{PhysRevLett.105.130401,1751-8121-44-49-492002,PhysRevE.85.051117}. Here, the notation $\mathscr{D}_{\alpha}\otimes\mathbbm{1}_{\overline{\alpha}}$ stands for some dissipator acting locally on qubit $\alpha$, in spite of using the full interacting Hamiltonian $H_{\text{ref}}=\sum_{\alpha}H_{0,\alpha}+H_I$ to account for the free dynamics.

\begin{figure}[t]
\includegraphics[width=8cm]{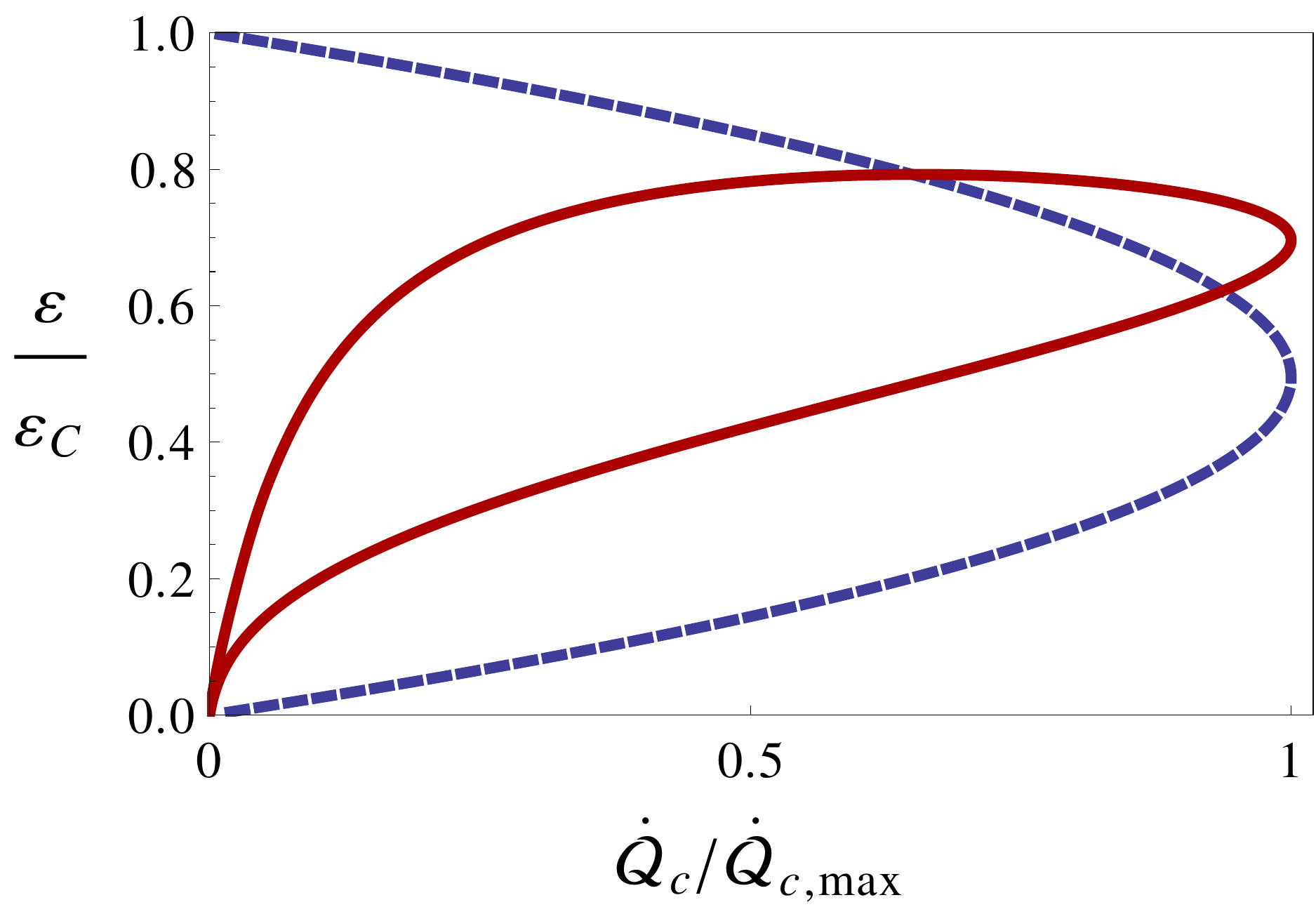}
\caption{(Color online). Comparison of the performance characteristics of the quantum absorption refrigerator according to the delocalized dissipative scheme of Eq.~\eqref{master} (solid), and to a ``localized'' master equation of the type Eq.~\eqref{masterlocal} (dashed) for the same choice of parameters (see text for details). All the quantities plotted are dimensionless.}
\label{fig2}
\end{figure}

Recall from the preceeding considerations, that such localized model for the dissipation is only physically consistent in the limit of strictly vanishing $g$ \cite{nicolaPriv}. On the contrary, if the realistic delocalized description of Eq.~\eqref{master} is adopted, given any value of the qubit-qubit interaction, no matter how small, it becomes impossible to approach $\varepsilon_C$ arbitrarily closely, as illustrated in Fig.~\ref{fig2}.
There we plot the \emph{performance caracteristics} for the refrigerator according to Eq.~\eqref{master} (solid), and as results from the localized master equation of the type Eq.~\eqref{masterlocal} (dashed) used in \cite{PhysRevLett.105.130401,1751-8121-44-49-492002,PhysRevE.85.051117}. In both cases, the operation temperatures are $T_h=66.25$ and $T_c=4.78$. We also fix $T_w=127.33$, $\omega_w=56.87$, $g=0.1$ and $\gamma=10^{-6}$ and $p_i=10^{-3}$. The only remaining free parameter, $\omega_c$, is varied from $0$ to the upper limit of the cooling window Eq.~\eqref{cwindow} $\omega_{c,\max}$, and the cooling power $\dot{\mathcal{Q}}_c$ for each configuration is plotted versus the corresponding COP, normalized by the Carnot bound $\varepsilon_C$. The cooling powers are also normalized by their maximum values $\dot{\mathcal{Q}}_{c,\max}$: $8.39\times 10^{-6}$ and $1.27\times 10^{-6}$ respectively. Clearly, $\omega_c=0$ corresponds to $\dot{\mathcal{Q}}_c=\varepsilon=0$. In the ideal case of Eq.~\eqref{masterlocal}, $\omega_c\rightarrow\omega_{c,\max}$ results in $\varepsilon\rightarrow\varepsilon_C$  and $\dot{\mathcal{Q}}_c\rightarrow 0$. However, any irreversibility would yield a COP not monotonically increasing with $\omega_c$, and therefore, a closed performance characteristic, as shown.

This suggests that the unavoidable delocalization in the dissipative dynamics makes the refrigerator non ideal and somehow wasteful, thus introducing a fundamental source of irreversibility that prevents it from cooling at the Carnot COP in practice.

\section{Performance bounds at maximum power}\label{secBound}
As we have just seen, in this three-qubit quantum absorption refrigerators, $\varepsilon_C$ cannot be approached \emph{arbitrarily closely} in practice. It is therefore crucial to introduce an alternative tight bound on some performance indicator, such that its saturation would mark the functioning of the refrigerator as effectively optimal. A sensible figure of merit in this context is, for instance, the COP $\varepsilon_*$ at maximum cooling power $\dot{\mathcal{Q}}_{c,\max}$. One can then seek to devise a general upper bound for such a performance indicator, and to characterize a region within the space of the control parameters $\{\omega_w,\,T_w\}$ that allows for its saturation. This would provide useful workpoints for the efficient implementation of the machine.

\begin{figure}[tb]
\includegraphics[width=7cm]{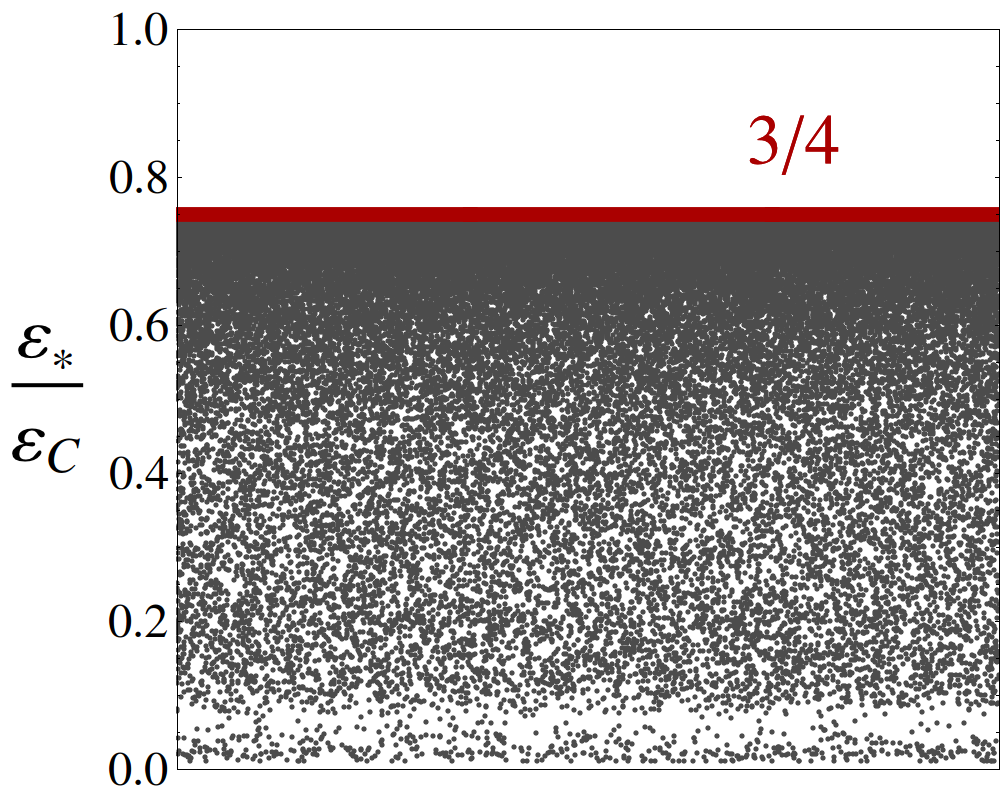}
\caption{(Color online). COP at maximum cooling power for $10^5$ random refrigerators, calculated from the stationary states of Eq.~\eqref{master}. The operation temperatures $T_c$ and $T_h$, as well as $T_w$, $\omega_w$, $g$ and $\gamma$, were chosen completely at random, always satisfying the constraints $k_B T_{\alpha}\gg \gamma$ (Born-Markov approximation) and $g\gg \gamma$ (rotating wave approximation). The value of $\omega_c$ yielding $\dot{\mathcal{Q}}_{c,\max}$ was found in each case (therefore fixing $\omega_h=\omega_c+\omega_w$) and the corresponding $\varepsilon_*/\varepsilon_C$, plotted. All the quantities plotted are dimensionless.}
\label{fig3}
\end{figure}

To investigate this issue, we run extensive numerics on the stationary states of Eq.~\eqref{master}, globally optimizing $\varepsilon_*$ over \emph{all} free parameters of the refrigerator, always under the consistency constraints implied by the Born-Markov and rotating wave approximations. We found that $\varepsilon_*$ is tightly upper bounded by
\begin{equation}\label{bondo}
\varepsilon_* \leq\frac{3}{4}\varepsilon_C.
\end{equation}
This is illustrated in Fig.~\ref{fig3}, where $\varepsilon_*/\varepsilon_C$ was plotted for $10^{5}$ quantum absorption refrigerators whose free parameters were all sampled at random.

In order to get analytical insight into the role and possible origin of the performance limit, we resort to the much simpler mathematical description based on the localized master equation of Refs.~\cite{PhysRevLett.105.130401,1751-8121-44-49-492002,PhysRevE.85.051117}. Even though the irreversibility associated with delocalized dissipation is completely absent in this ideal case, meaning that the COP can reach $\varepsilon_C$ (albeit at vanishing power), the COP at maximum cooling power is, nonetheless, still tightly upper bounded when optimization over all parameters is carried out. Specifically, a similar numerical analysis shows that the bound turns out to be $\frac12\varepsilon_C$ for the localized model. One can actually show this analytically, given $T_w$ and $\omega_w$ such that: (i) $\omega_{w}/T_{w,h}\ll 1$, and (ii) $\omega_{c,\text{max}}/T_c \ll 1$ (see Appendix \ref{secAppB} for a detailed proof), such a performance bound is saturated in the limit of \emph{large temperature difference} $T_c/T_h \ll 1$.

Interestingly, coming back now to the realistic situation modelled by Eq.~\eqref{master}, with a consistent treatment of delocalized dissipation, one sees that conditions (i) and (ii) are also sufficient to saturate the $\frac34 \varepsilon_C$ performance bound on the COP at maximum power when working at high difference between the operating temperatures. These conditions, therefore, provide the desired design prescriptions for the practical implementation of efficient quantum absorption refrigerators of this kind. However, it is in order to remark that those are just \emph{sufficient} conditions for optimal performance, and do not need to be necessarily met in order to attain a nearly optimal COP at maximum power: for instance, even the machine in Fig.~\ref{fig2} cools very close to the bound, despite having $\omega_w/T_h\sim 1$.

Finally, note that the fact that the performance bound differs quantitatively when the oversimplified localized picture is used instead ($\frac12$ vs $\frac34$ as a fraction of $\varepsilon_C$), should not be surprising, as the underlying dissipative dynamics also encloses essential differences.  The important point, however, is that the bound is also tight in that case, and that the analytical expression of the idealized stationary state is tractable enough and even proves insightful to obtain prescriptions for the saturation of the $\frac34 \varepsilon_C$ bound in the realistic model of Eq.~(\ref{master}).

\section{Stationary quantum correlations}\label{secQC}
We finally investigate the stationary bipartite quantum correlations established in the refrigerator, focusing on the realistic dissipative model of Eq.~(\ref{master}). Given the structure of $\varrho^{\infty}$,  the reduced states within the $2\times 2$ bipartitions $w\text{--}h$, $w\text{--}c$ and $h\text{--}c$ are diagonal and therefore, unentangled and completely classical. We then consider the only two-dimensional subspaces of $\bigotimes_\alpha\mathcal{H}_{S,\alpha}$ which are in direct interaction according to $H_\text{ref}$, namely those corresponding to the machine virtual qubit and the cold qubit. While no entanglement is found, more general quantum correlations measured by quantum discord \cite{olliver20011,henderson20011,rev} are \textit{always} present in this relevant bipartition \cite{ferraro}.
The structure of stationary quantum discord, analyzed in detail in the following,  does not however exhibit any specific relationship either with the maximization of $\dot{\mathcal{Q}}_c$ or $\varepsilon$, or with the behavior of $\varepsilon_*$ or $\dot{\mathcal{Q}}_{c,\max}$, as the control parameters $\{\omega_w,\,T_w\}$ are varied.
This supports the conclusion that the only essential quantum ingredient exploited by the machine is in fact the discreteness of its energy spectrum \cite{PhysRevLett.105.130401,1751-8121-44-49-492002,PhysRevE.85.051117}.

We recall that quantum discord is defined as
\begin{equation}D\left(\varrho_{AB}\right)\equiv\mathcal{I}\left(\varrho_{AB}\right)-\mathcal{I}(\sigma_{AB}),\end{equation} where the mutual information $\mathcal{I}(\varrho_{AB})$ quantifies total correlations in the bipartite state $\varrho_{AB}$, and $\sigma_{AB}$ is the post-measurement state after a minimally disturbing projective measurement on $B$.  We refer the reader to Refs.~\cite{olliver20011,henderson20011,rev} for details and interpretations of discord.

The dissipative dynamics of Eq.~\eqref{master} annihilates 3-qubit $X$-states that have $\varrho^{\infty}_{36}=\varrho^{\infty}_{63}$ as the only nonzero matrix elements outside the main diagonal when expressed in the computational basis. If any of the qubits is traced out, the remaining $2\times 2$ density matrix is diagonal and only involves the populations of $\varrho^{\infty}$. Therefore, entanglement and quantum discord in the bipartitions $w\text{--}h$, $w\text{--}c$ and $h\text{--}c$ vanish.

It is interesting to look instead to the only two-dimensional subspaces of $\bigotimes_{\alpha}\mathcal{H}_{S,\alpha}$ which are placed in direct interaction through $H_I$, that is, the machine virtual qubit and the cold qubit. The corresponding reduced state reads
\begin{gather*}
\rho^{\infty}_{vc}=\frac{P\varrho^{\infty} P}{\text{tr}P\varrho^{\infty} P},\\
\mbox{with }P \equiv \eeket{1}{0}{w}{h}\eebra{1}{0}{w}{h}+\eeket{0}{1}{w}{h}\eebra{0}{1}{w}{h}+\eket{0}{c}\ebra{0}{c}+\eket{1}{c}\ebra{1}{c}\,.
\end{gather*}
When expressed in the basis $\{\eeket{1}{0}{w}{h},\,\eeket{0}{1}{w}{h}\}\otimes\{\eket{0}{c},\,\eket{1}{c}\}$, the 2-qubit $X$-state $\rho^{\infty}_{vc}$ is given by
\begin{equation}
\varrho^{\infty}_{vc}=\frac{1}{\mathcal{N}}
	\left(
		\begin{array}{cccc}
			\varrho^{\infty}_{55} & 0 & 0 & 0 \\
			0 & \varrho^{\infty}_{66} & \varrho^{\infty}_{36} & 0 \\
			0 & \varrho^{\infty}_{36} & \varrho^{\infty}_{33} & 0 \\
			0 & 0 & 0 & \varrho^{\infty}_{44}
		\end{array}
	\right)\,,
\label{rhovc}\end{equation}
where $\mathcal{N}$ is the normalization factor. According to the positivity-of-the-partial-transpose separability criterion \cite{peres1996separability,horodecki1996separability}, the state $\varrho^{\infty}_{vc}$ is entangled iff
\begin{equation}
\varrho^{\infty}_{36}>\frac{1}{2}\frac{\varrho^{\infty}_{44}+\varrho^{\infty}_{55}}{\varrho^{\infty}_{44}-\varrho^{\infty}_{55}}\,.
\end{equation}

However, our stationary states are such that $\varrho^{\infty}_{36}\ll\varrho^{\infty}_{jj}$ for $j\in\{1,\cdots,8\}$, and therefore, no bipartite qubit entanglement can be present in them. On the contrary, one always finds nonzero stationary quantum discord between the machine virtual qubit and the cold qubit. In the search of the least disturbing local measurements for its quantification, we will restrict to projective measurements only. Since $\varrho^{\infty}_{vc}$ is an $X$-state, its discord can be computed analytically, using the formulas of Ref.~\cite{PhysRevA.81.042105}.

\begin{figure}[t]
\subfigure{
\includegraphics[width=4.2cm]{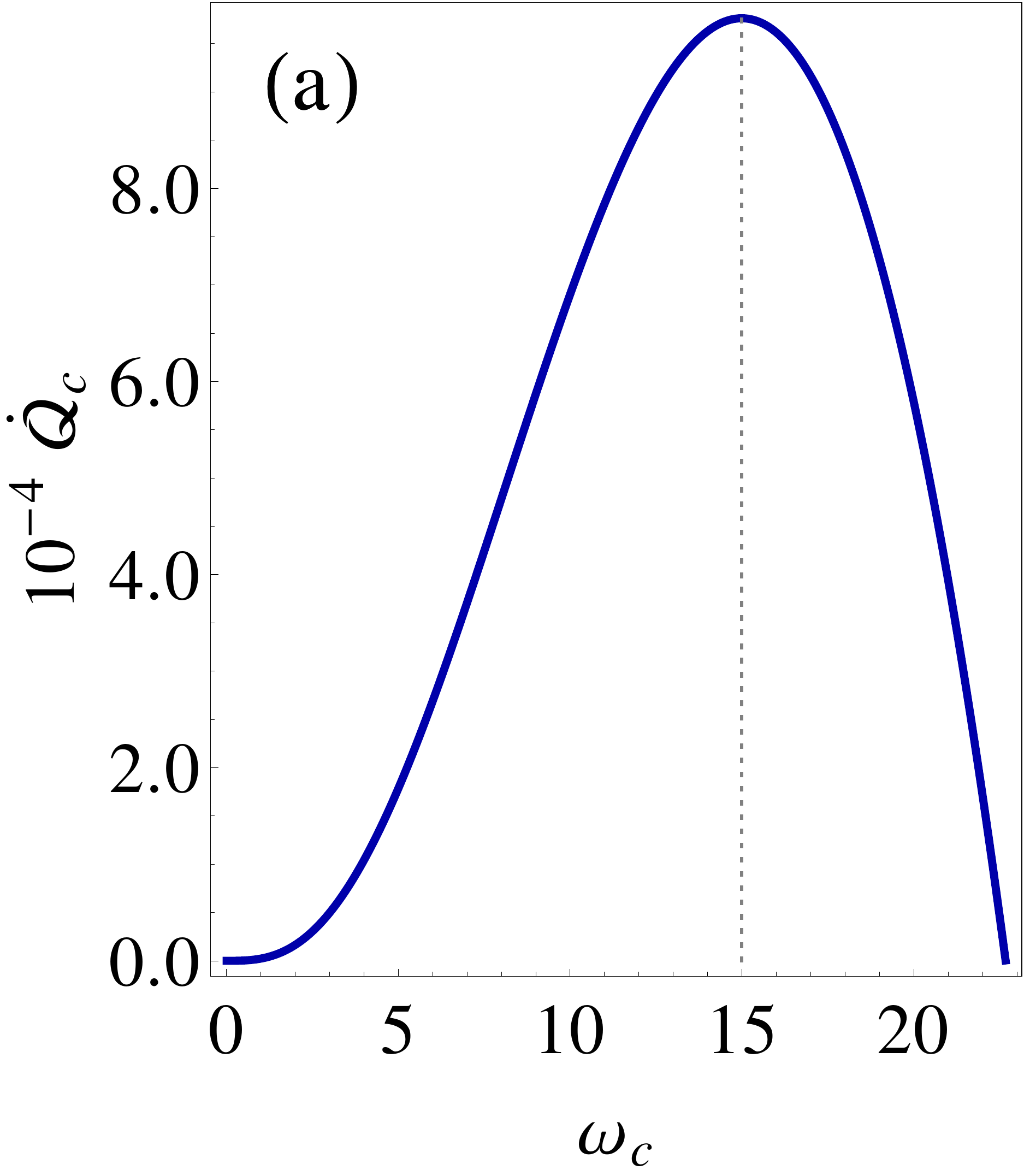}\label{fig2aS}}
\subfigure{
\includegraphics[width=4.2cm]{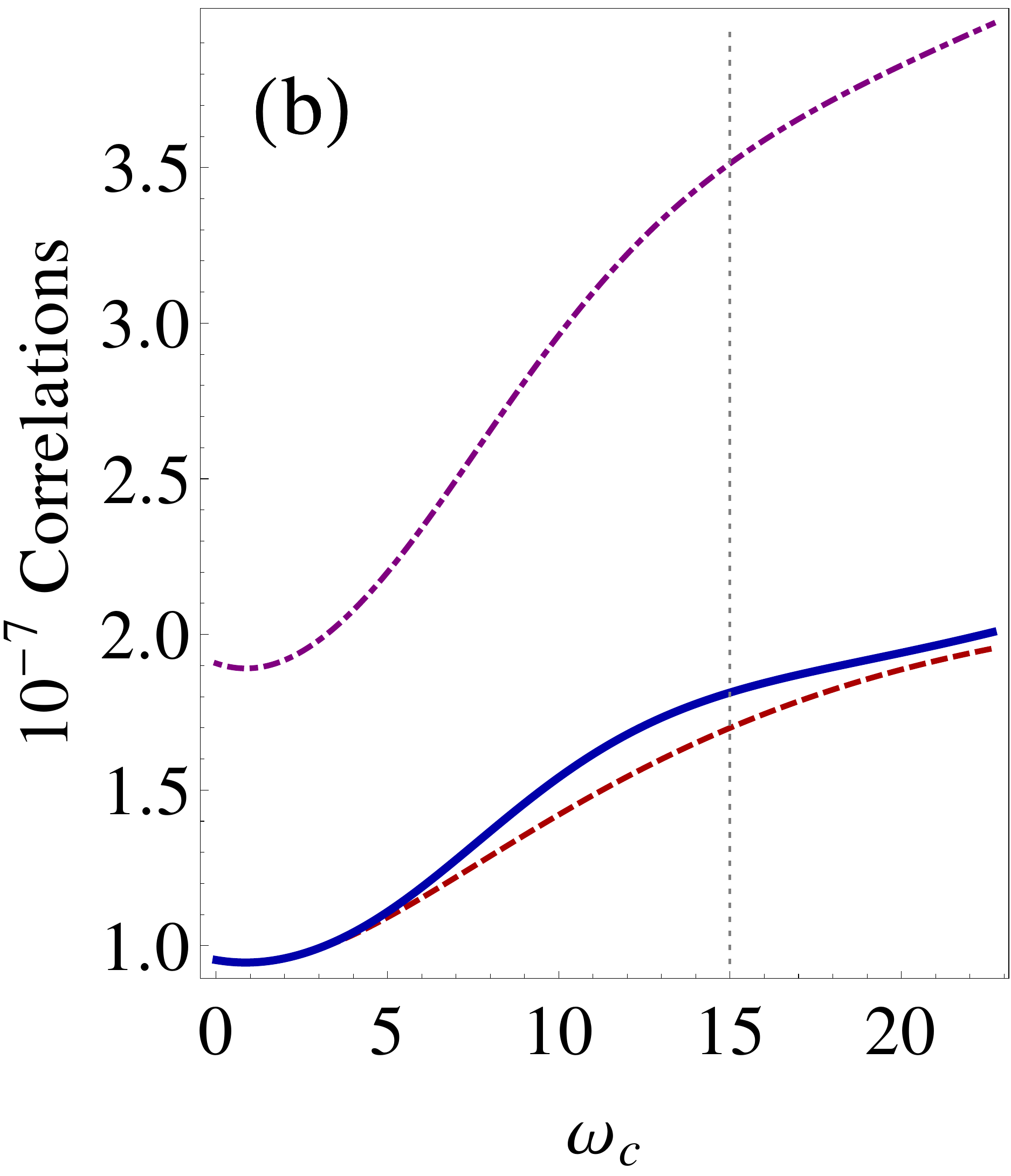}\label{fig2bS}}
\subfigure{
\includegraphics[width=4.2cm]{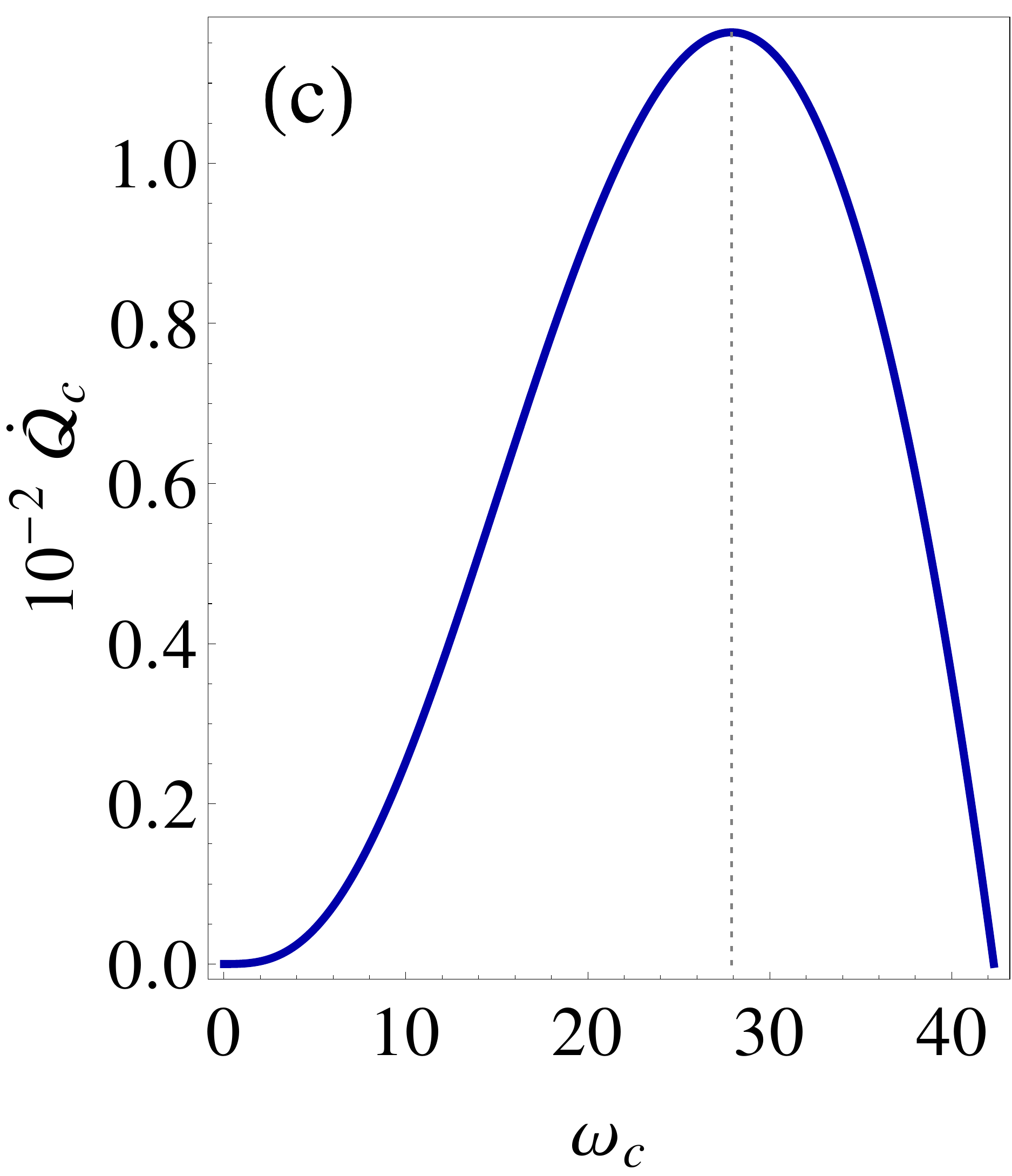}\label{fig2cS}}
\subfigure{
\includegraphics[width=4.2cm]{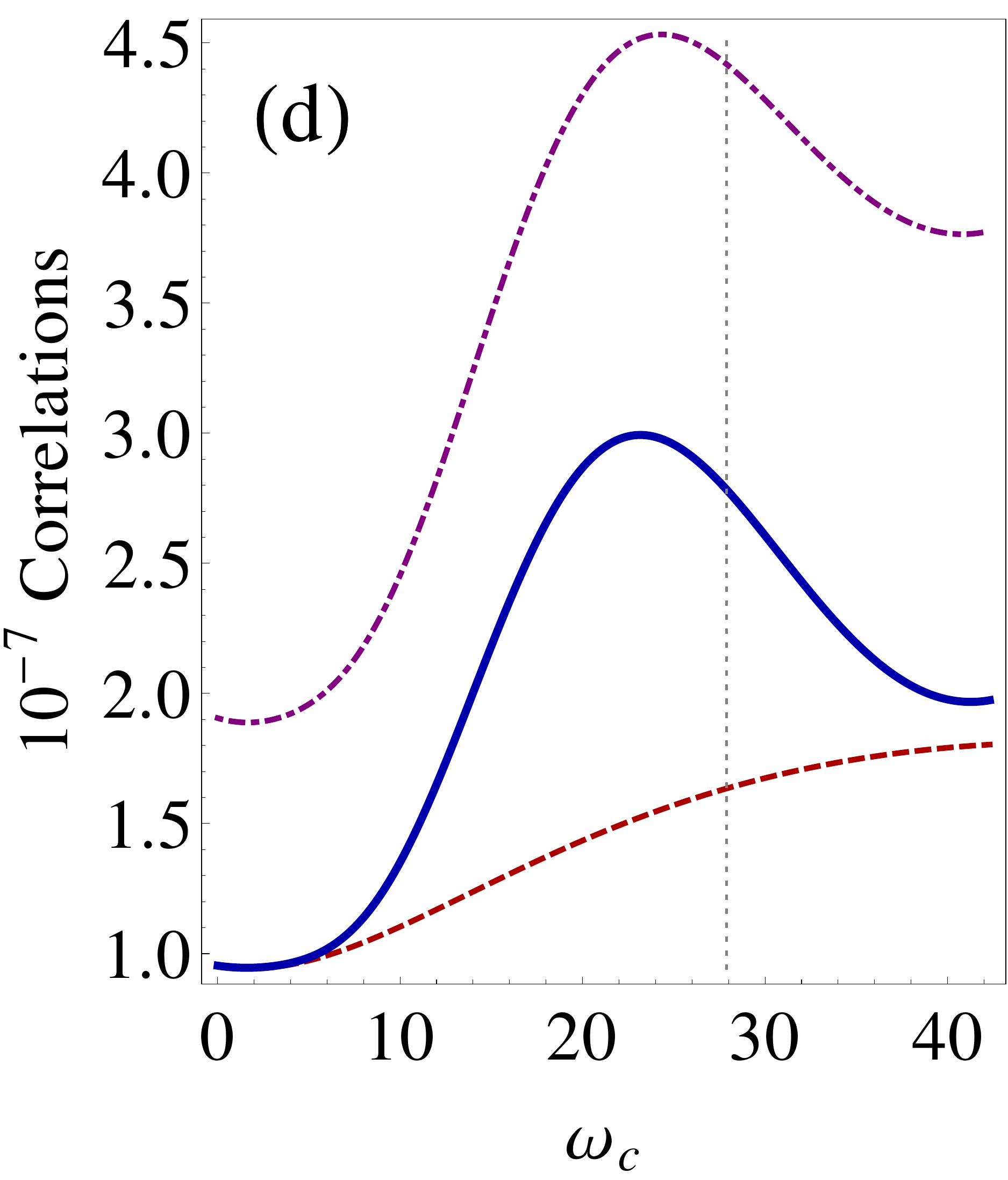}\label{fig2dS}}
\subfigure{
\includegraphics[width=4.2cm]{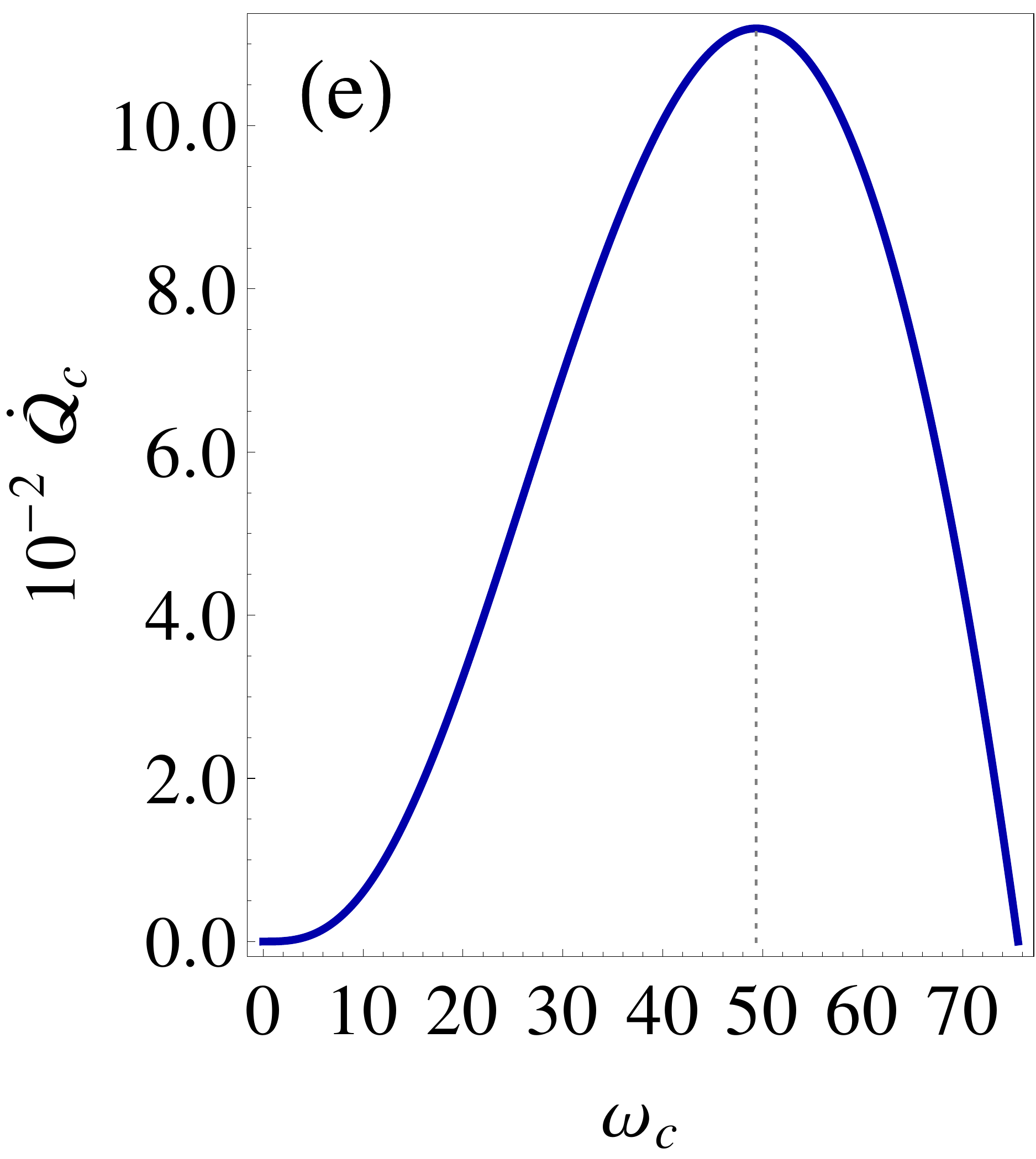}\label{fig2eS}}
\subfigure{
\includegraphics[width=4.2cm]{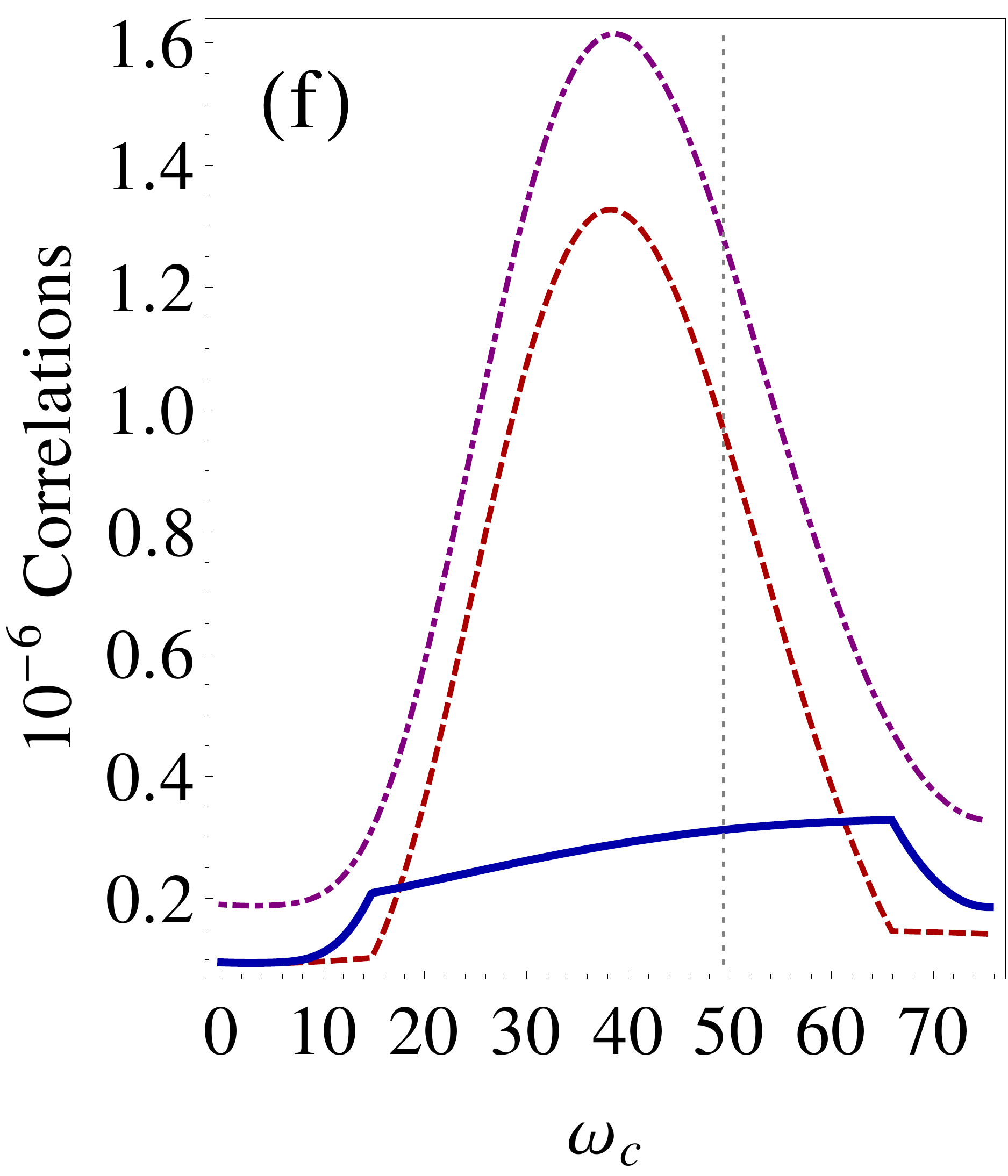}\label{fig2fS}}

\caption{(Left column) Cooling power $\dot{\mathcal{Q}}_c$ as a function of $\omega_c$ at fixed $T_\alpha$, $\omega_w$, $g$ and $\gamma$; and (Right column) Total (dot-dashed), classical (dashed) and quantum (solid) correlations for $T_w=180$, $T_h=95$, $T_c=80$, $g=0.1$, $\gamma=10^{-6}$ and $\omega_w=10$ [(a) and (b)], $\omega_w=15$ [(b) and (c)] and $\omega_w=30$ [(d) and (e)]. The gray dotted line marks the position of the frequency $\omega_{c,*}$ maximizing $\dot{\mathcal{Q}}_c$. All the quantities plotted are dimensionless.}
\label{fig2S}
\end{figure}

In the first place, we fixed $T_\alpha$ and $\omega_w$, and looked at the stationary total [$\mathcal{I}(\varrho_{vc})$], quantum [$D(\varrho_{vc})$] and classical [$\mathcal{I}(\sigma_{vc})$] correlations between the machine virtual qubit and the cold qubit, to see whether they play any role in the maximization of the cooling power $\dot{\mathcal{Q}}_c$ for $0<\omega_c\leq \omega_{c,\max}$ (see Fig.~\ref{fig2S}). Similarly to $\dot{\mathcal{Q}}_c$, for \emph{intermediate} values of $\omega_w$ all correlations are peaked around a certain value of $\omega_c$ that nevertheless usually differs from the frequency $\omega_{c,*}$ that maximizes $\dot{\mathcal{Q}}_c$ [cf. Figs.~\ref{fig2cS} and  \ref{fig2dS}]. Smaller work frequencies yield a monotonic behaviour of correlations instead, as shown in Fig.~\ref{fig2bS}, while larger values of $\omega_w$ reveal a more intricate structure [see Fig.~\ref{fig2fS}].

In Figs.~\ref{fig2bS} and \ref{fig2dS}, the measurements that maximize $\mathcal{I}(\sigma_{vc})$ consist in e.g. projections onto $\eket{+}{c}\equiv\left(\eket{0}{c}+\eket{1}{c}\right)/\sqrt{2}$ and $\eket{-}{c}\equiv\left(\eket{0}{c}-\eket{1}{c}\right)/\sqrt{2}$ for any $\omega_c$. However, in Fig.~\ref{fig2fS} projective measurements in the computational basis of the cold qubit, become optimal in the interval $14\lesssim\omega_c\lesssim 66$. These discontinuous changes in the optimal measurement schemes result in a non differentiable classical correlations and quantum discord. In all three cases, the COP increases linearly with $\omega_c$ and starts to decrease only as $\omega_{c,\max}$ is approached. It seems therefore, clear that the maximization of $\dot{\mathcal{Q}}_c$ and $\varepsilon$ are not related in any way to the only non vanishing $2\times 2$ stationary quantum correlations in the system.

\begin{figure}[t]
\subfigure{
\includegraphics[width=4.2cm]{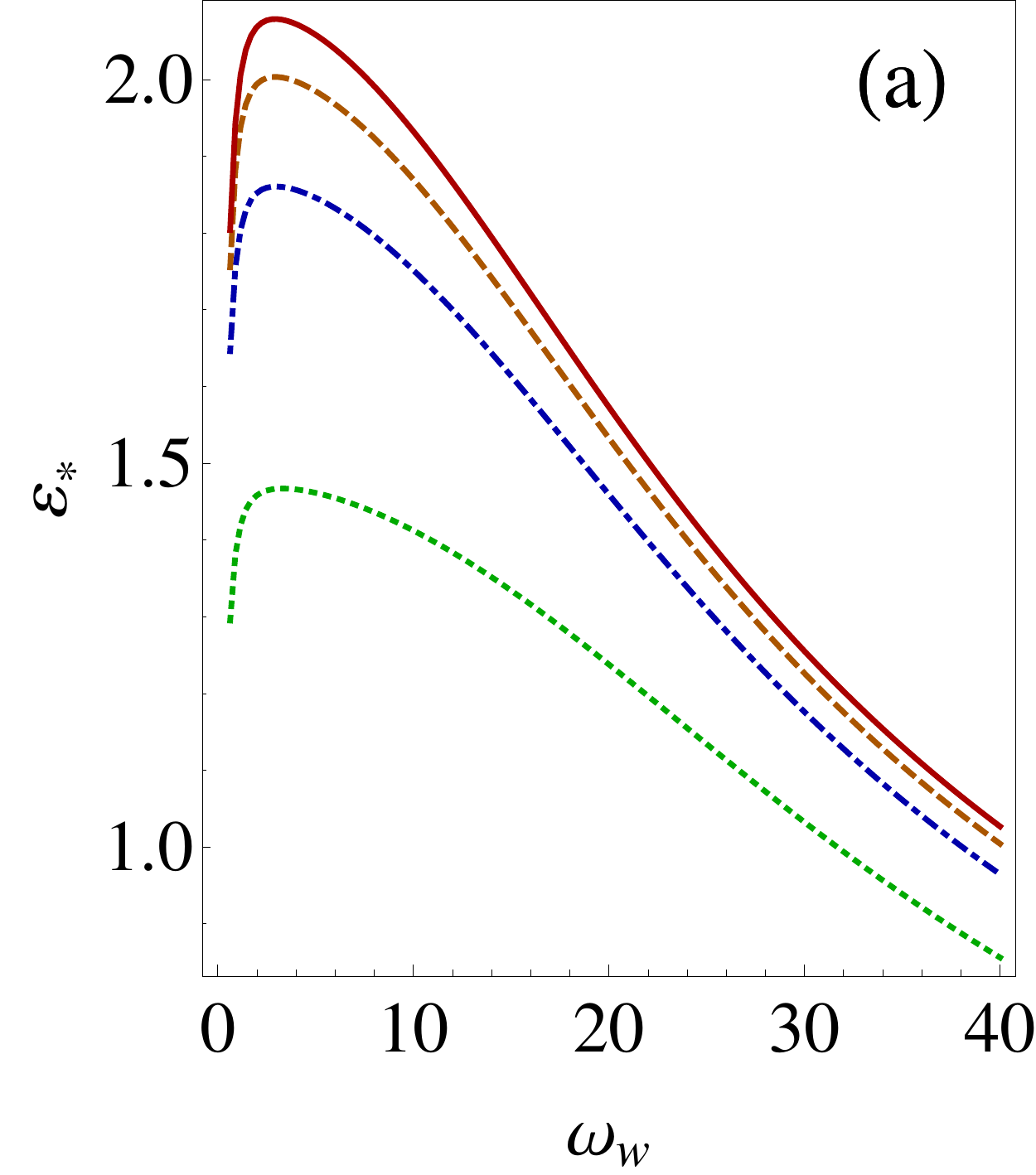}\label{fig3aS}}
\subfigure{
\includegraphics[width=4.2cm]{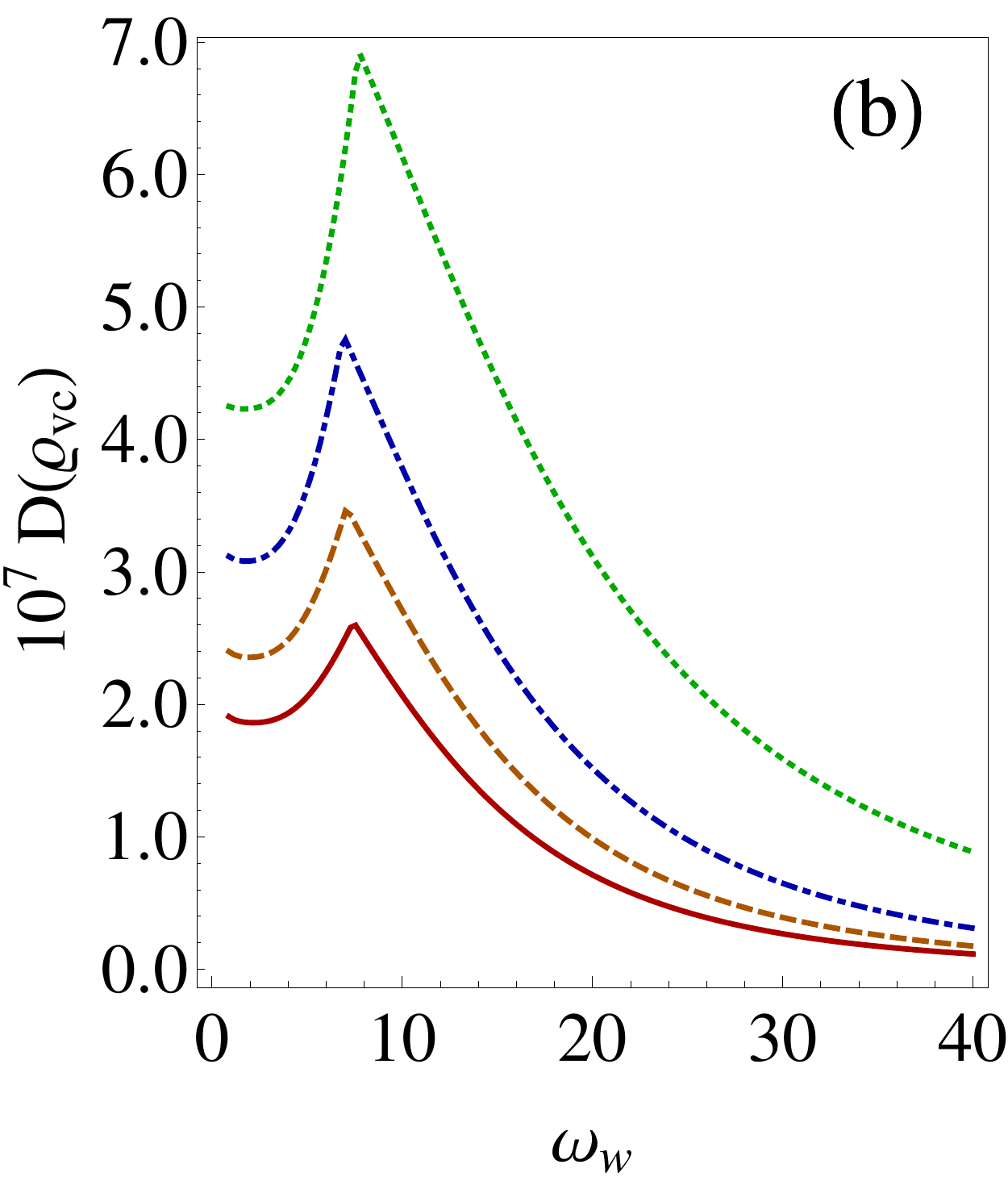}\label{fig3bS}}

\caption{(a) COP at maximum cooling power $\varepsilon_*$ and (b) quantum discord at $\omega_{c,*}$ as a function of $\omega_w$ for different temperatures $T_w$. Parameters were chosen as $T_h=17$, $T_c=13$, $g=0.1$ and $\gamma=2.5\times 10^{-5}$ and $T_w=50$ (dotted), $T_w=100$ (dot-dashed), $T_w=150$ (dashed) and $T_w=200$ (solid). All the quantities plotted are dimensionless.}
\label{fig3S}
\end{figure}

We also ruled out any possible interplay between quantum discord at $\omega_{c,*}$ and the maximization of $\varepsilon_*$ and $\dot{\mathcal{Q}}_{c,\max}$, using $\omega_w$ and $T_w$ as control parameters. In Fig.~\ref{fig3S} we plot $\varepsilon_*$ and the corresponding quantum discord as a function of $\omega_w$ for different temperatures $T_w$. After increasing abruptly for small work frequencies, the COP at maximum power starts to decay as $\omega_w$ grows. When it comes to its temperature dependence, $\varepsilon_*$ seems to increase with $T_w$ as shown in Fig.~\ref{fig3S}. In Fig.~\ref{fig3bS}, we see that for small $\omega_w$, quantum discord is also an increasing function of the work frequency. The optimal measurement scheme changes from $\{\eket{+}{c}\ebra{+}{c},\,\eket{-}{c}\ebra{-}{c}\}$ to $\{\eket{0}{c}\ebra{0}{c},\,\eket{1}{c}\ebra{1}{c}\}$ at some $\tilde{\omega}_w$, which produces a sharp maximum. For $\omega_w>\tilde{\omega}_w$, discord decays monotonically. At any fixed $\omega_w$, it is decreasing with the work temperature. The corresponding $\dot{\mathcal{Q}}_{c,\max}$ grows exponentially with $\omega_w$ and also increases as a function of $T_w$.

Even if the maximum discord at fixed $T_w$ does not coincide with the maximum of $\varepsilon_*$, it still marks a useful operation point of the refrigerator where a certain compromise between $\dot{\mathcal{Q}}_{c,\max}$ and $\varepsilon_*$ is achieved. Our results also suggest that the COP at maximum power and the corresponding cooling power are enhanced, at fixed $\omega_w$ (and sufficiently small $g$), at the expense of the destruction of quantum correlations. It is possible, however, to increase these two figures of merit and yet build more quantumness in the system if one also leaves $\omega_w$ as a free parameter.

From all the preceeding we see that there is no clear relationship between the quantumness of the bipartite qubit-qubit correlations established in the stationary regime and the efficient performance of the refrigerator. In the idealized case of vanishing $g$, the ability of this quantum machine to saturate the Carnot COP comes from the discreteness of its energy spectrum, in contrast with any \emph{continuous} ``classical'' counterpart \cite{PhysRevE.85.051117}. Actually this discreteness and thermalization are the only essential building blocks for the basic operation of the refrigerator. It may not be surprising then, that any residual quantumness of correlations appears in the system as a by-product rather than as fundamental resource for its enhanced performance.

\section{Conclusions}\label{secConcl}
We consistently studied the three-qubit quantum absorption chillers introduced in \cite{PhysRevLett.105.130401,1751-8121-44-49-492002,PhysRevE.85.051117} adopting a physically meaningful system-bath interaction model. The resulting delocalized dissipation effects prevent the refrigerator {\it a priori} from cooling arbitrarily closely to the Carnot COP $\varepsilon_C$, thus introducing unavoidable irreversibility in the stationary cooling process.

As an alternative to $\varepsilon_C$, a more useful performance bound had to be considered instead to assess the optimality of a given realization of such thermal devices. We chose to look at the COP $\varepsilon_*$ at maximum cooling power $\dot{\mathcal{Q}}_{c,\max}$, and found that global optimization over all model parameters yields a tight upper bound on $\varepsilon_*$ of $\frac34 \varepsilon_C$. Sufficient conditions to saturate it in the limit of large temperature difference were also given.

The efficient performance of these machines was not found to relate in any obvious way to stationary bipartite total, classical or quantum stationary correlations present in the system.

Understanding the role played by the correlation properties of the environment on the performance limit could render the practical prescriptions for the realization of even more efficient quantum refrigerators, accessible with present-day technology \cite{0295-5075_97_4_40003, 1210.3649} and will warrant further investigation. The extension of our results to a wider range of quantum absorption chillers will also be a subject of a future dedicated study.

\acknowledgments{
The authors are grateful to D. Girolami, N. Lo Gullo, K. Hovhannisyan, R. Kosloff, J. Goold, A. Ac\'{i}n, M. Perarnau, R. Vasile, R. Gallego, D. Cavalcanti, M. Navascu\'{e}s and M. Huber for fruitful discussions and constructive criticism. This project was funded by the Spanish MICINN (Grant No. FIS2010-19998) and the European Union (FEDER), by the Canary Islands Government through the ACIISI fellowships (85\% co funded by European Social Fund), and by the University of Nottingham through an Early Career Research and Knowledge Transfer Award and an EPSRC Research Development Fund Grant (PP-0313/36).}

\appendix

\section{Derivation of the Markovian master equation}\label{secAppA}
We will now consistently build the Markovian master equation Eq.~(\ref{master}) for the reduced state $\varrho$ of the three refrigerator qubits.

The process starts by taking the interaction picture with respect to the \emph{free Hamiltonian} $H_F=\sum_\alpha H_{0,\alpha}+H_I+\sum_\alpha H_{B,\alpha}$. An initial preparation, uncorrelated between system and environment is chosen so that $\rho\left(0\right)=\varrho\left(0\right)\otimes\chi$, where $\chi\equiv\bigotimes_{\alpha}\mathcal{Z}^{-1}_{\alpha}e^{-H_{B,\alpha}/T_{\alpha}}$. This choice guarantees on the one hand that the reduced evolution is a completely positive (and trace preserving) dynamical map (CPT) \cite{PhysRevLett.102.100402} and on the other, that the average of the bath operator $B_\alpha\equiv\sum_\lambda g_\lambda(a_{\alpha,\lambda}-a^{\dagger}_{\alpha,\lambda})$ vanishes initially $\text{tr}_B B_\alpha \rho\left(0\right)=0$, and actually also at any latter time as long as the Born approximation holds (see below).

The interaction picture Liouville-Von Neumann equation for $\rho\left(t\right)$ is then suitably manipulated. Next, the Born or weak dissipation approximation, according to which $\rho\left(t\right)\simeq\varrho\left(t\right)\otimes\chi$ is performed. The Markov approximation, that consists in neglecting any memory effects in the reduced evolution, finally leads to (see \cite{breuer2002theory} for details)
\begin{equation}
\dot{\tilde{\varrho}}=-\sum_{\alpha,\,\beta}\int_0^{\infty}ds\,\text{tr}_{B}\left[H_{D,\,\alpha}\left(t\right),\left[H_{D,\,\beta}\left(t-s\right),\tilde{\varrho}\left(t\right)\otimes\chi\right]\right]\,,
\label{markov0}\end{equation}
where $\tilde{\varrho}\left(t\right)\equiv e^{iH_Ft}\varrho\left(t\right)e^{-iH_Ft}$ stands for the interaction picture reduced state of the refrigerator qubits, and $H_{D,\alpha}\left(t\right)\equiv e^{iH_Ft}H_{D,\alpha}e^{-iH_Ft}$. The dynamical map $\Phi(t,0)$ evolving $\tilde{\varrho}\left(0\right)$ into $\tilde{\varrho}\left(t\right)$ that results from Eq.~\eqref{markov0}, has the \emph{semigroup property} under map composition: $\Phi\left(t,0\right)\cdot\Phi\left(s,0\right)=\Phi\left(t+s,0\right)$, which implies that Eq.~\eqref{markov0} can be cast in the standard Lindblad form \cite{lindblad1976generators} of Eq.~\eqref{master}.

In order to achieve this, we shall decompose the system operators $\sigma_{x_\alpha}$ from the system-baths interaction term $H_I=\sum_\alpha \sigma_{x_\alpha}\otimes B_\alpha$ into eigen-operators $\mathcal{A}_{\alpha,\omega}$ of $H_\text{ref}$ such that
\begin{equation}
\sigma_{x_{\alpha}}=\sum_{\omega}\mathcal{A}_{\alpha,\omega}\,,\qquad \left[H_{\text{ref}},\mathcal{A}_{\alpha,\omega}\right]=-\omega \mathcal{A}_{\alpha,\omega} \,.
\end{equation}

The non-Hermitian Lindblad or jump operators $\mathcal{A}_{\alpha,\omega}$ are defined as
\begin{equation}
\mathcal{A}_{\alpha,\omega}=\sum'_{\omega_k-\omega_j=\omega}\ket{j}\bra{j}\sigma_{x_{\alpha}}\ket{k}\bra{k},
\label{jumpoper}\end{equation}
where $\ket{j}$ is an eigenstate of $H_\text{ref}$ with energy $\omega_j$ and non-degeneracy is assumed. The eigenvalues of $H_\text{ref}$ are $\{0,\,\omega_w,\,2\omega_h,\,\omega_c,\,\omega_w+\omega_h,\,\omega_h+\omega_c,\,\omega_h-g,\,\omega_h+g\}$, and their corresponding eigenvectors
\begin{align*}
\ket{1}&=\eeeket{0}{0}{0}{w}{h}{c},\, \ket{2}=\eeeket{1}{0}{0}{w}{h}{c},\, \ket{3}=\eeeket{1}{1}{1}{w}{h}{c},\\
\ket{4}&=\eeeket{0}{0}{1}{w}{h}{c},\,\ket{5}=\eeeket{1}{1}{0}{w}{h}{c},\, \ket{6}=\eeeket{0}{1}{1}{w}{h}{c}, \\
\ket{7}&=\left(\eeeket{1}{0}{1}{w}{h}{c} - \eeeket{0}{1}{0}{w}{h}{c}\right)/\sqrt{2}\,, \\
\ket{8}&=\left(\eeeket{1}{0}{1}{w}{h}{c} + \eeeket{0}{1}{0}{w}{h}{c}\right)/\sqrt{2}\,.
\end{align*}

Therefore, from Eq.~\eqref{jumpoper} it is easy to see that there are only six open \emph{decay channels} (i.e. transition frequencies $\omega=\omega_j-\omega_k$ with non-zero $\mathcal{A}_{\alpha,\omega}$) for each bath $\alpha$, corresponding to $\{\pm\omega_\alpha,\,\pm\omega_{\alpha}\pm g\}$. These jump operators are explicitly
\begin{align*}
\mathcal{A}_{w,\,\omega_w}&=\sqrt{\gamma}\left(\ket{1}\bra{2}+\ket{6}\bra{3}\right),\\
\mathcal{A}_{w,\,\omega_w+g}&=\sqrt{\gamma}\left(\ket{4}\bra{8}-\ket{7}\bra{5}\right)/\sqrt{2},\\
\mathcal{A}_{w,\,\omega_w-g}&=\sqrt{\gamma}\left(\ket{4}\bra{7}+\ket{8}\bra{5}\right)/\sqrt{2},\\
\\
\mathcal{A}_{h,\,\omega_h}&=\sqrt{\gamma}\left(\ket{2}\bra{5}+\ket{4}\bra{6}\right),\\
\mathcal{A}_{h,\,\omega_h+g}&=\sqrt{\gamma}\left(\ket{7}\bra{3}+\ket{1}\bra{8}\right)/\sqrt{2},\\
\mathcal{A}_{h,\,\omega_h-g}&=\sqrt{\gamma}\left(\ket{8}\bra{3}-\ket{1}\bra{7}\right)/\sqrt{2},\\
\\
\mathcal{A}_{c,\,\omega_c}&=\sqrt{\gamma}\left(\ket{1}\bra{4}+\ket{5}\bra{3}\right),\\
\mathcal{A}_{c,\,\omega_c+g}&=\sqrt{\gamma}\left(\ket{2}\bra{8}-\ket{7}\bra{6}\right)/\sqrt{2},\\
\mathcal{A}_{c,\,\omega_c-g}&=\sqrt{\gamma}\left(\ket{2}\bra{7}+\ket{8}\bra{6}\right)/\sqrt{2}\,.
\end{align*}

The remaining Lindblad operators are just given by the adjoint of these, since $\mathcal{A}^{\dagger}_{\alpha,\omega}=\mathcal{A}_{\alpha,-\omega}$.

The interaction picture system-environment coupling Hamiltonian may be now written as
\begin{equation}
H_I\left(t\right)=e^{iH_Ft}H_Ie^{-iH_Ft}=\sum_{\alpha,\omega}e^{-i\omega t}\mathcal{A}_{\alpha,\omega}\otimes B_{\alpha}\left(t\right),
\end{equation}
where $B_{\alpha}\left(t\right)=\sum_\lambda g_\lambda (a_\lambda e^{-i\omega_\lambda t}-a^{\dagger}_\lambda e^{i\omega_\lambda t})$. Gathering all this back into Eq.~\eqref{markov0} yields
\begin{equation}
\dot{\tilde{\varrho}}=\sum_{\alpha,\,\omega,\,\omega'} e^ {i(\omega'-\omega)t}C_{\alpha,\omega}\left(\mathcal{A}_{\alpha,\omega}\tilde{\varrho} \mathcal{A}^{\dagger}_{\alpha,\omega'}-\mathcal{A}^{\dagger}_{\alpha,\omega'}\mathcal{A}_{\alpha,\omega}\tilde{\varrho}\right)+\text{h.c.}\,,
\label{markov1}\end{equation}
where the (complex) bath correlations $C_{\alpha,\,\omega}$ are defined as
\begin{equation}
C_{\alpha,\omega}=\int_0^\infty ds\, e^{i\omega s}\,\text{tr}_B \,\chi B_\alpha\left(t\right)B_\alpha\left(t-s\right)\equiv \frac{1}{2}\Gamma_{\alpha,\,\omega}+i S_{\alpha,\,\omega}\,.
\end{equation}

Note that due to our choice of the initial preparation and provided that the Born approximation holds, all bath correlations depending on $\text{tr}_B\,\chi B_\alpha\left(t\right)B_\beta\left(t-s\right)$ with $\alpha\neq\beta$ vanish, i.e. the baths are independent.

The details of $C_{\alpha,\omega}$ may be also worked out, exploiting that the baths were prepared in a thermal equilibrium state \cite{breuer2002theory}. Choosing $g_\lambda\propto\sqrt{\omega_{\lambda}}$, the spectral correlation tensor $\Gamma_{\alpha,\omega}$ reads
\begin{equation}
\Gamma_{\alpha,\omega}\propto\omega^3 e^{\beta\omega/2}\left(\sinh\frac{\beta\omega}{2} \right)^{-1},
\label{spectr}\end{equation}
where the proportionality constant is of order one and may be absorbed into $\gamma$.

Another important step towards the derivation of the master equation Eq.~\eqref{master} is the assumption that the typical system time-scales $|\omega-\omega'|^{-1}$ with $\omega\neq\omega'$, are much smaller than the relaxation time $\gamma^{-1}$, which allows to discard all rapidly oscillating terms $\omega\neq\omega'$ that average to zero in a coarse-grained picture of the reduced dynamics (rotating wave approximation). This leaves (we refer again to \cite{breuer2002theory} for details)
\begin{equation}
\dot{\tilde{\varrho}}=\sum_{\alpha,\omega} \Gamma_{\alpha,\omega} \left( \mathcal{A}_{\alpha,\omega}\tilde{\varrho} \mathcal{A}^{\dagger}_{\alpha,\omega} -\frac{1}{2} \{ \mathcal{A}^{\dagger}_{\alpha,\omega} \mathcal{A}_{\alpha,\omega}, \tilde{\varrho} \}_+\right)\,,
\label{masterint}\end{equation}
where we have discarded the Lamb-Shift term $-i\sum_{\alpha,\omega}S_{\alpha,\omega}\left[\mathcal{A}^{\dagger}_{\alpha,\omega}\mathcal{A}_{\alpha,\omega},\tilde{\varrho}\right]$ as usually done when working in the quantum optical regime.

The only thing that remains to be done in order to recover Eq.~\eqref{master}, is to transform Eq.~\eqref{masterint} back into de Schr\"{o}dinger picture by noting that
\begin{equation}
\dot{\varrho}=-iH_F e^{-iH_Ft}\tilde{\varrho}e^{iH_Ft}+ie^{-iH_Ft}\tilde{\varrho}e^{iH_Ft}H_F+e^{-iH_Ft}\dot{\tilde{\varrho}}e^{iH_Ft}\,.
\end{equation}

This immediately yields Eq.~\eqref{master} if one identifies $A_{\alpha,\omega}$ with $e^{-iH_Ft}\mathcal{A}_{\alpha,\omega}e^{iH_Ft}=e^{i\omega t}\mathcal{A}_{\alpha,\omega}$.

The reduced dynamics generated by equation Eq.~\eqref{master} may be understood as a stochastic process in the Hilbert state space of the refrigerator qubits, in which the deterministic evolution of any pure state is interrupted by discontinuous quantum jumps $\ket{\psi}\mapsto \mathcal{N}^{-1}A_{\alpha,\omega}\ket{\psi}$, occurring at rates $\Gamma_{\alpha,\omega}$. The density matrix $\varrho\left(t\right)$ at any time $t$ is recovered as an esemble average over stochastic trajectories \cite{osa_93,breuer2002theory}.

It is clear that all interaction picture jump operators $\mathcal{A}_{\alpha,\omega}$ introduced above, can produce delocalized dissipation (quantum jumps) in the sense discussed in the main article. As $g$ gets closer to zero, the rates $\Gamma_{\alpha,\pm\omega_{\alpha}+g}$ and $\Gamma_{\alpha,\pm\omega_{\alpha}-g}$ linearly approach each other, meaning that the jump processes $\mathcal{A}_{\alpha,\pm\omega+g}$ and $\mathcal{A}_{\alpha,\pm\omega-g}$ become equally likely, so that their delocalized contributions start to compensate. In the strict limit of $g=0$, only one transition frequency ($\omega_\alpha$) is left for each bath, the corresponding Lindblad operator being just the sum $\mathcal{A}_{\alpha,\pm\omega}+\mathcal{A}_{\alpha,\pm\omega+g}+\mathcal{A}_{\alpha,\pm\omega-g}\propto\sigma_{\alpha,\mp}$ [cf. Eq.~\eqref{jumpoper}], which is a localized jump operator. Note that even if the difference $|\Gamma_{\alpha,\pm\omega+g}-\Gamma_{\alpha,\pm\omega-g}|$ depends linearly on $g\ll 1$, the exponentials in Eq.~\eqref{spectr} make it extremely sensitive to the qubit-qubit interaction strength, so that even a slight departure from the non interacting case can make delocalized dissipation effects very important.

Let us finally comment on the underlying assumptions leading to Eq.~\eqref{master}. Even though the rotating wave approximation makes the problem much more tractable, it is not essential and one could just avoid it [cf. Eq.~\eqref{markov1}]. Situations in which the dissipation times become comparable to the system time scale (i.e. the realm of quantum Brownian motion) may be then accounted for, conceivably resulting in qualitative differences. Accounting for the non negligible renormalization effects of the system-environment interaction on the system itself, becomes important in these cases. On the contrary, within the quantum optical regime, the rotating wave approximation only slightly modifies the reduced dynamics but not the stationary states of the refrigerator, thus leaving our results unaffected.

If the baths are assumed to have some structure, the Markov assumption cannot be consistently performed. Nevertheless, as long as the dissipation strength remains sufficiently weak so that the Born approximation is still in place, no qualitative differences should be expected from what we report in the main article. Finally, if the dissipation becomes strong enough, the thermalization of a single isolated qubit in contact with its corresponding bath is no longer guaranteed, so that the basic operation of the refrigerator is compromised.

\section{Analytical derivation of the performance bound $\frac12\varepsilon_C$ under localized dissipation}\label{secAppB}

Considering the localized dissipative model of \cite{PhysRevLett.105.130401,1751-8121-44-49-492002,PhysRevE.85.051117}, we shall now prove that its COP at maximum power $\varepsilon_*$ is upper bounded by $\frac12\varepsilon_C$, whenever conditions
\begin{subequations}\label{conditions}
\begin{align}
\omega_{w} &\ll T_{w,h} \tag{i}\label{i}\\
\omega_{w} &\ll \tau\equiv \frac{T_w(T_h-T_c)}{T_w-T_h} \tag{ii} \label{ii}
\end{align}
\end{subequations}
are met, at large temperature difference $T_c/T_h\ll 1$. \newline

\medskip
\noindent {\it Proof. } Our starting point will be Eqs.~(18) and (8)-(10) in Ref.\cite{1751-8121-44-49-492002}, where the cooling power $\dot{\mathcal{Q}}_c$ was given as
\begin{equation}
\dot{\mathcal{Q}}_{c,w}= q \frac{\Delta}{2+\frac{q^2}{2g^2}+\sum_\alpha q_\alpha +\sum_{\alpha\beta}Q_{\alpha\beta}\Omega_{\alpha\beta}}\omega_{c,w}\tag{\theequation} \,,
\label{cpower}\end{equation}
and
\begin{subequations}
\begin{align}
\Delta &= \frac{e^{-(\omega_w + \omega_c)/T_h}-e^{\omega_w/T_w}e^{-\omega_c/T_c}}{(1+e^{\omega_w/T_w})(1+e^{(\omega_w + \omega_c)/T_h})(1+e^{\omega_c /T_c})} \label{delta}\\
\Omega_{\alpha\beta} &= \left\{
     \begin{array}{lr}
       \frac{1}{1+e^{-\omega_\alpha/T_\alpha}}\frac{e^{-\omega_\beta/T_\beta}}{1+e^{-\omega_\beta/T_\beta}}+\frac{e^{-\omega_\alpha/T_\alpha}}{1+e^{-\omega_\alpha/T_\alpha}}\frac{1}{1+e^{-\omega_\beta/T_\beta}}	 \qquad (\alpha,\,\beta \neq h)\\
       \frac{e^{-\omega_\alpha/T_\alpha}}{1+e^{-\omega_\alpha/T_\alpha}}\frac{e^{-\omega_\beta/T_\beta}}{1+e^{-\omega_\beta/T_\beta}}+\frac{1}{1+e^{-\omega_\alpha/T_\alpha}}\frac{1}{1+e^{-\omega_\beta/T_\beta}}	 \qquad (\beta\neq\alpha = h) \\
       \frac{1}{1+e^{-\omega_\alpha/T_\alpha}}\frac{1}{1+e^{-\omega_\beta/T_\beta}}+\frac{e^{-\omega_\alpha/T_\alpha}}{1+e^{-\omega_\alpha/T_\alpha}}\frac{1}{1+e^{-\omega_\beta/T_\beta}}	 \qquad (\alpha \neq\beta = h) \\
       \frac{e^{-\omega_\alpha/T_\alpha}}{1+e^{-\omega_\alpha/T_\alpha}}\frac{1}{1+e^{-\omega_\beta/T_\beta}}+\frac{1}{1+e^{-\omega_\alpha/T_\alpha}}\frac{1}{1+e^{-\omega_\beta/T_\beta}}	 \qquad (\alpha=\beta= h)\,.
     \end{array}
   \right. \label{omega}
\end{align}
\end{subequations}

Here, $q$, $q_i$ and $Q_{\alpha\beta}$ only depend on the three, possibly different, dissipation rates $p_i$, while $\Delta$ and $\Omega_{\alpha\beta}$ depend on all temperatures and frequencies. All we need to do is to find the $\omega_{c,*}$ that maximizes Eq.~\eqref{cpower}, and then compute the corresponding $\varepsilon_*$.

First of all, note that conditions \eqref{i} and \eqref{ii} imply $\omega_c/T_c\ll 1$ for any $\omega_c<\omega_{c,\max}$, i.e. within the cooling window
\begin{equation}
1>e^{-\omega_c / T_c}>e^{-\omega_{c,\max}/ T_c}>e^{-\omega_w/\tau}\simeq 1-\frac{\omega_w}{T_w},
\end{equation}
Due to the ordering $T_w>T_h>T_c$ in the bath's equilibrium temperatures, we must also have $\omega_c/T_h\ll 1$. Since $\omega_h=\omega_w+\omega_c$, this translates into $\omega_h/T_h\ll 1$. Therefore, conditions \eqref{i} and \eqref{ii} can be alternatively stated as
\begin{equation}
\omega_{w,h,c}\ll T_{w,h,c}.
\end{equation}
Furthermore, in the limit of large temperature difference $T_c/T_h \ll 1$, assuming that $T_c/T_h$ is at least of order $\omega_w/T_h$ and $\omega_w/\tau$, we also have
\begin{equation}
\frac{\omega_c}{T_h}\leq \frac{\omega_{c,\max}}{T_h}=\frac{\omega_w}{\tau}\frac{T_c}{T_h}\ll \frac{\omega_w}{T_h},
\end{equation}
As a consequence, Eqs.~\eqref{delta} and \eqref{omega} can be expressed as
\begin{subequations}\label{app}
\begin{align}
\Delta &\simeq \frac{1}{8} \left(e^{-\omega_w /T_h}-e^{\omega_w/T_w}e^{-\omega_c/T_c}\right) + \mathcal{O}\left(\frac{\omega_\alpha}{T_\alpha}\right) \label{detalapp}\\
\Omega_{\alpha\beta} &\simeq \frac{1}{2} + \mathcal{O}\left(\frac{\omega_\alpha}{T_\alpha}\right)\, . \label{omegaapp}
\end{align}
\end{subequations}
In this regime, the denominator of $\dot{\mathcal{Q}}_c$ as given in Eq.~\eqref{cpower} becomes independent of $\omega_c$, so that its maximization is equivalent to that of Eq.~\eqref{detalapp}. This yields
\begin{equation}
\left(1-\frac{\omega_{c,*}}{T_c}\right)e^{\left(1-\omega_{c,*}/T_c\right)}=e^{(\omega_w /T_w-\omega_w/T_h)}.
\end{equation}

The solution to an equation of the form $x\,e^x=a$ may be expressed in terms of the \emph{Lambert-W function} or \emph{product-logarithm} \cite{LambertW}, as $x=W_0\left(a\, e\right)$. Therefore $\omega_{c,*}$ reads
\begin{equation}
\omega_{c,*} = T_c\left[1-W_0\left(e^{(1+\omega_w /T_w-\omega_w/T_h)}\right)\right] \, .
\label{omegac0}\end{equation}
Among the properties of $W_0\left(z\right)$, we shall make use of its series expansion around $z=e$
\begin{equation}
W_0\left(z\right)=\frac{1}{2}+\frac{z}{2e}+...
\label{lambertexp}\end{equation}
Taking again \eqref{i} into account, $e^{(1+\omega_w /T_w-\omega_w/T_h)}\simeq 1+\omega_w /T_w-\omega_w/T_h$ which, combined with Eqs.~\eqref{omegac0} and \eqref{lambertexp}, results in
\begin{equation}
\omega_{c,*} \simeq \frac{\omega_w T_c}{2}\left(\frac{1}{T_h} -\frac{1}{T_w}\right) \, .
\end{equation}
The COP at maximum cooling power $\varepsilon_*=\omega_{c,*}/\omega_w$ \cite{1751-8121-44-49-492002} normalized by $\varepsilon_C$, may be thus approximated by
\begin{equation}
\frac{\varepsilon_*}{\varepsilon_C}\simeq \frac{T_c}{2\varepsilon_C}\left(\frac{1}{T_h} -\frac{1}{T_w}\right)=\frac{1}{2}\left(1 -\frac{T_c}{T_h}\right)\leq\frac{1}{2} \,,
\end{equation}
which saturates in the limit of large temperature difference $T_c/T_h\ll 1$.\hfill $\Box$

%As already stated in the main text, the fact that the performance bound is numerically different under the inconsistent localized description of dissipation is of no physical relevance. What is essential is that design prescriptions (i) and (ii) also lead to saturation of the $\frac34 \varepsilon_C$ bound at large temperature difference $T_c/T_h\ll 1$ when the dissipation is addressed consistently.

\bibliographystyle{apsrev}
%\bibliography{/home/luis/Escritorio/Current/10/References}

\begin{thebibliography}{20}
\expandafter\ifx\csname natexlab\endcsname\relax\def\natexlab#1{#1}\fi
\expandafter\ifx\csname bibnamefont\endcsname\relax
  \def\bibnamefont#1{#1}\fi
\expandafter\ifx\csname bibfnamefont\endcsname\relax
  \def\bibfnamefont#1{#1}\fi
\expandafter\ifx\csname citenamefont\endcsname\relax
  \def\citenamefont#1{#1}\fi
\expandafter\ifx\csname url\endcsname\relax
  \def\url#1{\texttt{#1}}\fi
\expandafter\ifx\csname urlprefix\endcsname\relax\def\urlprefix{URL }\fi
\providecommand{\bibinfo}[2]{#2}
\providecommand{\eprint}[2][]{\url{#2}}

\bibitem[{\citenamefont{Geva and Kosloff}(1996)}]{geva1996quantum}
\bibinfo{author}{\bibfnamefont{E.}~\bibnamefont{Geva}} \bibnamefont{and}
  \bibinfo{author}{\bibfnamefont{R.}~\bibnamefont{Kosloff}},
  \bibinfo{journal}{J. Chem. Phys.} \textbf{\bibinfo{volume}{104}},
  \bibinfo{pages}{7681} (\bibinfo{year}{1996}).

\bibitem[{\citenamefont{Palao et~al.}(2001)\citenamefont{Palao, Kosloff, and
  Gordon}}]{PhysRevE.64.056130}
\bibinfo{author}{\bibfnamefont{J.~P.} \bibnamefont{Palao}},
  \bibinfo{author}{\bibfnamefont{R.}~\bibnamefont{Kosloff}}, \bibnamefont{and}
  \bibinfo{author}{\bibfnamefont{J.~M.} \bibnamefont{Gordon}},
  \bibinfo{journal}{Phys. Rev. E} \textbf{\bibinfo{volume}{64}},
  \bibinfo{pages}{056130} (\bibinfo{year}{2001}).

\bibitem[{\citenamefont{Levy and Kosloff}(2012)}]{PhysRevLett.108.070604}
\bibinfo{author}{\bibfnamefont{A.}~\bibnamefont{Levy}} \bibnamefont{and}
  \bibinfo{author}{\bibfnamefont{R.}~\bibnamefont{Kosloff}},
  \bibinfo{journal}{Phys. Rev. Lett.} \textbf{\bibinfo{volume}{108}},
  \bibinfo{pages}{070604} (\bibinfo{year}{2012}).

\bibitem[{\citenamefont{Levy et~al.}(2012)\citenamefont{Levy, Alicki, and
  Kosloff}}]{PhysRevE.85.061126}
\bibinfo{author}{\bibfnamefont{A.}~\bibnamefont{Levy}},
  \bibinfo{author}{\bibfnamefont{R.}~\bibnamefont{Alicki}}, \bibnamefont{and}
  \bibinfo{author}{\bibfnamefont{R.}~\bibnamefont{Kosloff}},
  \bibinfo{journal}{Phys. Rev. E} \textbf{\bibinfo{volume}{85}},
  \bibinfo{pages}{061126} (\bibinfo{year}{2012}).

\bibitem[{\citenamefont{Linden et~al.}(2010)\citenamefont{Linden, Popescu, and
  Skrzypczyk}}]{PhysRevLett.105.130401}
\bibinfo{author}{\bibfnamefont{N.}~\bibnamefont{Linden}},
  \bibinfo{author}{\bibfnamefont{S.}~\bibnamefont{Popescu}}, \bibnamefont{and}
  \bibinfo{author}{\bibfnamefont{P.}~\bibnamefont{Skrzypczyk}},
  \bibinfo{journal}{Phys. Rev. Lett.} \textbf{\bibinfo{volume}{105}},
  \bibinfo{pages}{130401} (\bibinfo{year}{2010}).

\bibitem[{\citenamefont{Skrzypczyk et~al.}(2011)\citenamefont{Skrzypczyk,
  Brunner, Linden, and Popescu}}]{1751-8121-44-49-492002}
\bibinfo{author}{\bibfnamefont{P.}~\bibnamefont{Skrzypczyk}},
  \bibinfo{author}{\bibfnamefont{N.}~\bibnamefont{Brunner}},
  \bibinfo{author}{\bibfnamefont{N.}~\bibnamefont{Linden}}, \bibnamefont{and}
  \bibinfo{author}{\bibfnamefont{S.}~\bibnamefont{Popescu}},
  \bibinfo{journal}{J. Phys. A: Math. Theor.} \textbf{\bibinfo{volume}{44}},
  \bibinfo{pages}{492002} (\bibinfo{year}{2011}).

\bibitem[{\citenamefont{Brunner et~al.}(2012)\citenamefont{Brunner, Linden,
  Popescu, and Skrzypczyk}}]{PhysRevE.85.051117}
\bibinfo{author}{\bibfnamefont{N.}~\bibnamefont{Brunner}},
  \bibinfo{author}{\bibfnamefont{N.}~\bibnamefont{Linden}},
  \bibinfo{author}{\bibfnamefont{S.}~\bibnamefont{Popescu}}, \bibnamefont{and}
  \bibinfo{author}{\bibfnamefont{P.}~\bibnamefont{Skrzypczyk}},
  \bibinfo{journal}{Phys. Rev. E} \textbf{\bibinfo{volume}{85}},
  \bibinfo{pages}{051117} (\bibinfo{year}{2012}).

\bibitem[{\citenamefont{Zhou and Segal}(2010)}]{PhysRevE.82.011120}
\bibinfo{author}{\bibfnamefont{Y.}~\bibnamefont{Zhou}} \bibnamefont{and}
  \bibinfo{author}{\bibfnamefont{D.}~\bibnamefont{Segal}},
  \bibinfo{journal}{Phys. Rev. E} \textbf{\bibinfo{volume}{82}},
  \bibinfo{pages}{011120} (\bibinfo{year}{2010}).

\bibitem[{\citenamefont{Gelbwaser-Klimovsky
  et~al.}(2012)\citenamefont{Gelbwaser-Klimovsky, Alicki, and
  Kurizki}}]{1209.1190}
\bibinfo{author}{\bibfnamefont{D.}~\bibnamefont{Gelbwaser-Klimovsky}},
  \bibinfo{author}{\bibfnamefont{R.}~\bibnamefont{Alicki}}, \bibnamefont{and}
  \bibinfo{author}{\bibfnamefont{G.}~\bibnamefont{Kurizki}},
  \bibinfo{journal}{e-print arXiv:1209.1190}  (\bibinfo{year}{2012}).

\bibitem{natpop}
S. Popescu, A. J. Short,  and A. Winter, Nature Phys. {\bf 2}, 754 (2006).

\bibitem{guld}
R. Dorner, J. Goold, C. Cormick, M. Paternostro, and V. Vedral,
Phys. Rev. Lett. {\bf 109}, 160601 (2012)

\bibitem{renner}
D. Egloff, O. C. O. Dahlsten, R. Renner, and V. Vedral, e-print arXiv:1207.0434 (2012).

\bibitem{giazotto}
 F. Giazotto, T. T. Heikkil\"a, A. Luukanen, A. M. Savin,
and J. P. Pekola, Rev. Mod. Phys. {\bf 78}, 217 (2006).

\bibitem[{\citenamefont{Chen and Li}(2012)}]{0295-5075_97_4_40003}
\bibinfo{author}{\bibfnamefont{Y.-X.} \bibnamefont{Chen}} \bibnamefont{and}
  \bibinfo{author}{\bibfnamefont{S.-W.} \bibnamefont{Li}},
  \bibinfo{journal}{Europhys. Lett.} \textbf{\bibinfo{volume}{97}},
  \bibinfo{pages}{40003} (\bibinfo{year}{2012}).

\bibitem[{\citenamefont{Venturelli et~al.}(2012)\citenamefont{Venturelli,
  Fazio, and Giovannetti}}]{1210.3649}
\bibinfo{author}{\bibfnamefont{D.}~\bibnamefont{Venturelli}},
  \bibinfo{author}{\bibfnamefont{R.}~\bibnamefont{Fazio}}, \bibnamefont{and}
  \bibinfo{author}{\bibfnamefont{V.}~\bibnamefont{Giovannetti}}
  (\bibinfo{year}{2012}), \bibinfo{note}{e-print arXiv:1210.3649}.

\bibitem[{\citenamefont{Abah et~al.}(2012)\citenamefont{Abah, Ro\ss{}nagel,
  Jacob, Deffner, Schmidt-Kaler, Singer, and Lutz}}]{PhysRevLett.109.203006}
\bibinfo{author}{\bibfnamefont{O.}~\bibnamefont{Abah}},
  \bibinfo{author}{\bibfnamefont{J.}~\bibnamefont{Ro\ss{}nagel}},
  \bibinfo{author}{\bibfnamefont{G.}~\bibnamefont{Jacob}},
  \bibinfo{author}{\bibfnamefont{S.}~\bibnamefont{Deffner}},
  \bibinfo{author}{\bibfnamefont{F.}~\bibnamefont{Schmidt-Kaler}},
  \bibinfo{author}{\bibfnamefont{K.}~\bibnamefont{Singer}}, \bibnamefont{and}
  \bibinfo{author}{\bibfnamefont{E.}~\bibnamefont{Lutz}},
  \bibinfo{journal}{Phys. Rev. Lett.} \textbf{\bibinfo{volume}{109}},
  \bibinfo{pages}{203006} (\bibinfo{year}{2012}).

\bibitem[{\citenamefont{Curzon and Ahlborn}(1975)}]{curzon1975efficiency}
\bibinfo{author}{\bibfnamefont{F.}~\bibnamefont{Curzon}} \bibnamefont{and}
  \bibinfo{author}{\bibfnamefont{B.}~\bibnamefont{Ahlborn}},
  \bibinfo{journal}{Am. J. Phys.} \textbf{\bibinfo{volume}{43}},
  \bibinfo{pages}{22} (\bibinfo{year}{1975}).

\bibitem[{\citenamefont{Velasco et~al.}(1997)\citenamefont{Velasco, Roco,
  Medina, and Hern\'andez}}]{PhysRevLett.78.3241}
\bibinfo{author}{\bibfnamefont{S.}~\bibnamefont{Velasco}},
  \bibinfo{author}{\bibfnamefont{J.~M.~M.} \bibnamefont{Roco}},
  \bibinfo{author}{\bibfnamefont{A.}~\bibnamefont{Medina}}, \bibnamefont{and}
  \bibinfo{author}{\bibfnamefont{A.~C.} \bibnamefont{Hern\'andez}},
  \bibinfo{journal}{Phys. Rev. Lett.} \textbf{\bibinfo{volume}{78}},
  \bibinfo{pages}{3241} (\bibinfo{year}{1997}).

\bibitem[{\citenamefont{Esposito et~al.}(2010)\citenamefont{Esposito, Kawai,
  Lindenberg, and Van~den Broeck}}]{PhysRevLett.105.150603}
\bibinfo{author}{\bibfnamefont{M.}~\bibnamefont{Esposito}},
  \bibinfo{author}{\bibfnamefont{R.}~\bibnamefont{Kawai}},
  \bibinfo{author}{\bibfnamefont{K.}~\bibnamefont{Lindenberg}},
  \bibnamefont{and} \bibinfo{author}{\bibfnamefont{C.}~\bibnamefont{Van~den
  Broeck}}, \bibinfo{journal}{Phys. Rev. Lett.} \textbf{\bibinfo{volume}{105}},
  \bibinfo{pages}{150603} (\bibinfo{year}{2010}).

\bibitem[{\citenamefont{Wang et~al.}(2012)\citenamefont{Wang, Li, Tu,
  Hern\'andez, and Roco}}]{PhysRevE.86.011127}
\bibinfo{author}{\bibfnamefont{Y.}~\bibnamefont{Wang}},
  \bibinfo{author}{\bibfnamefont{M.}~\bibnamefont{Li}},
  \bibinfo{author}{\bibfnamefont{Z.~C.} \bibnamefont{Tu}},
  \bibinfo{author}{\bibfnamefont{A.~C.} \bibnamefont{Hern\'andez}},
  \bibnamefont{and} \bibinfo{author}{\bibfnamefont{J.~M.~M.}
  \bibnamefont{Roco}}, \bibinfo{journal}{Phys. Rev. E}
  \textbf{\bibinfo{volume}{86}}, \bibinfo{pages}{011127}
  (\bibinfo{year}{2012}).

\bibitem[{\citenamefont{Allahverdyan et~al.}(2010)\citenamefont{Allahverdyan,
  Hovhannisyan, and Mahler}}]{PhysRevE.81.051129}
\bibinfo{author}{\bibfnamefont{A.~E.} \bibnamefont{Allahverdyan}},
  \bibinfo{author}{\bibfnamefont{K.}~\bibnamefont{Hovhannisyan}},
  \bibnamefont{and} \bibinfo{author}{\bibfnamefont{G.}~\bibnamefont{Mahler}},
  \bibinfo{journal}{Phys. Rev. E} \textbf{\bibinfo{volume}{81}},
  \bibinfo{pages}{051129} (\bibinfo{year}{2010}).

\bibitem[{\citenamefont{Ollivier and Zurek}(2001)}]{olliver20011}
\bibinfo{author}{\bibfnamefont{H.}~\bibnamefont{Ollivier}} \bibnamefont{and}
  \bibinfo{author}{\bibfnamefont{W.~H.} \bibnamefont{Zurek}},
  \bibinfo{journal}{Phys. Rev. Lett.} \textbf{\bibinfo{volume}{88}},
  \bibinfo{pages}{017901} (\bibinfo{year}{2001}).

\bibitem[{\citenamefont{Henderson and Vedral}(2001)}]{henderson20011}
\bibinfo{author}{\bibfnamefont{L.}~\bibnamefont{Henderson}} \bibnamefont{and}
  \bibinfo{author}{\bibfnamefont{V.}~\bibnamefont{Vedral}},
  \bibinfo{journal}{J. of Phys. A: Math. Gen.} \textbf{\bibinfo{volume}{34}},
  \bibinfo{pages}{6899} (\bibinfo{year}{2001}).

\bibitem[{\citenamefont{Breuer and Petruccione}(2002)}]{breuer2002theory}
\bibinfo{author}{\bibfnamefont{H.}~\bibnamefont{Breuer}} \bibnamefont{and}
  \bibinfo{author}{\bibfnamefont{F.}~\bibnamefont{Petruccione}},
  \emph{\bibinfo{title}{The Theory of Open Quantum Systems}}
  (\bibinfo{publisher}{Oxford University Press, USA}, \bibinfo{year}{2002}).



%\bibitem[{Not({\natexlab{a}})}]{Note1}
%\bibinfo{note}{Here, delocalized dissipation should not be
%  thought of as referring to memory effects in the dissipative
%  dynamics.}



\bibitem[{\citenamefont{Lo~Gullo}()}]{nicolaPriv}
\bibinfo{author}{\bibfnamefont{N.}~\bibnamefont{Lo~Gullo}},
  \bibinfo{note}{(private communication).}



\bibitem{rev} K. Modi, A. Brodutch, H. Cable, T. Paterek, and V. Vedral, Rev. Mod. Phys. {\bf 84}, 1655 (2012).

\bibitem{ferraro} A. Ferraro, L. Aolita, D. Cavalcanti, F. M. Cucchietti, and A. Ac\'in, Phys. Rev. A {\bf 81}, 052318 (2010).



\bibitem[{\citenamefont{Peres}(1996)}]{peres1996separability}
\bibinfo{author}{\bibfnamefont{A.}~\bibnamefont{Peres}},
  \bibinfo{journal}{Phys. Rev. Lett.} \textbf{\bibinfo{volume}{77}},
  \bibinfo{pages}{1413} (\bibinfo{year}{1996}).

\bibitem[{\citenamefont{Horodecki et~al.}(1996)\citenamefont{Horodecki,
  Horodecki, and Horodecki}}]{horodecki1996separability}
\bibinfo{author}{\bibfnamefont{M.}~\bibnamefont{Horodecki}},
  \bibinfo{author}{\bibfnamefont{P.}~\bibnamefont{Horodecki}},
  \bibnamefont{and}
  \bibinfo{author}{\bibfnamefont{R.}~\bibnamefont{Horodecki}},
  \bibinfo{journal}{Phys. Lett. A} \textbf{\bibinfo{volume}{223}},
  \bibinfo{pages}{1} (\bibinfo{year}{1996}).

\bibitem[{\citenamefont{Ali et~al.}(2010)\citenamefont{Ali, Rau, and
  Alber}}]{PhysRevA.81.042105}
\bibinfo{author}{\bibfnamefont{M.}~\bibnamefont{Ali}},
  \bibinfo{author}{\bibfnamefont{A.~R.~P.} \bibnamefont{Rau}},
  \bibnamefont{and} \bibinfo{author}{\bibfnamefont{G.}~\bibnamefont{Alber}},
  \bibinfo{journal}{Phys. Rev. A} \textbf{\bibinfo{volume}{81}},
  \bibinfo{pages}{042105} (\bibinfo{year}{2010}).

\bibitem[{\citenamefont{Shabani and Lidar}(2009)}]{PhysRevLett.102.100402}
\bibinfo{author}{\bibfnamefont{A.}~\bibnamefont{Shabani}} \bibnamefont{and}
  \bibinfo{author}{\bibfnamefont{D.~A.} \bibnamefont{Lidar}},
  \bibinfo{journal}{Phys. Rev. Lett.} \textbf{\bibinfo{volume}{102}},
  \bibinfo{pages}{100402} (\bibinfo{year}{2009}).


\bibitem[{\citenamefont{Lindblad}(1976)}]{lindblad1976generators}
\bibinfo{author}{\bibfnamefont{G.}~\bibnamefont{Lindblad}},
  \bibinfo{journal}{Comm. Math. Phys.} \textbf{\bibinfo{volume}{48}},
  \bibinfo{pages}{119} (\bibinfo{year}{1976}).

\bibitem[{\citenamefont{Molmer et~al.}(1993)\citenamefont{Molmer, Castin, and
  Dalibard}}]{osa_93}
\bibinfo{author}{\bibfnamefont{K.}~\bibnamefont{Molmer}},
  \bibinfo{author}{\bibfnamefont{Y.}~\bibnamefont{Castin}}, \bibnamefont{and}
  \bibinfo{author}{\bibfnamefont{J.}~\bibnamefont{Dalibard}},
  \bibinfo{journal}{Journal of the Optical Society of America B}
  \textbf{\bibinfo{volume}{10}}, \bibinfo{pages}{524} (\bibinfo{year}{1993}).


\bibitem[{\citenamefont{Corless et~al.}(1996)\citenamefont{Corless, Gonnet,
  Hare, Jeffrey, and Knuth}}]{LambertW}
\bibinfo{author}{\bibfnamefont{R.~M.} \bibnamefont{Corless}},
  \bibinfo{author}{\bibfnamefont{G.~H.} \bibnamefont{Gonnet}},
  \bibinfo{author}{\bibfnamefont{D.~E.~G.} \bibnamefont{Hare}},
  \bibinfo{author}{\bibfnamefont{D.~J.} \bibnamefont{Jeffrey}},
  \bibnamefont{and} \bibinfo{author}{\bibfnamefont{D.~E.} \bibnamefont{Knuth}},
  \bibinfo{journal}{Adv. Comput. Math.} \textbf{\bibinfo{volume}{5}},
  \bibinfo{pages}{329} (\bibinfo{year}{1996}).


\end{thebibliography}

\end{document}